\documentclass[aps,pra,twocolumn,showpacs,superscriptaddress]{revtex4-1}
\usepackage{graphicx}
\usepackage{epstopdf}
\usepackage{amsmath}
\usepackage{amssymb}
\usepackage{float}
\usepackage{bm}
\usepackage{mathtools}
\usepackage{multirow}
\usepackage{color}

\renewcommand{\u}[1]{\underline{#1}}

\input{epsf}
\begin{document}

%\title{Collective coordinate separability of fractional quantum Hall states}
\title{Hyperspherical theory of the quantum Hall effect:
  the role of exceptional degeneracy}

\author{K.~M. Daily}
\affiliation{Department of Physics and Astronomy,
Purdue University,
West Lafayette, Indiana 47907, USA}
\author{R.~E. Wooten}
\affiliation{Department of Physics and Astronomy,
Purdue University,
West Lafayette, Indiana 47907, USA}
\author{Chris H. Greene}
\affiliation{Department of Physics and Astronomy,
Purdue University,
West Lafayette, Indiana 47907, USA}

\date{\today}

\begin{abstract}
By separating the Schr\"odinger equation for $N$ noninteracting spin-polarized 
fermions in two-dimensional hyperspherical coordinates, we demonstrate 
that fractional quantum Hall (FQH) states emerge naturally from 
degeneracy patterns of the antisymmetric free-particle eigenfunctions. 
In the presence of Coulomb interactions, the FQH states split 
off from a degenerate manifold and become observable as distinct 
quantized energy eigenstates with an energy gap. This alternative 
classification scheme is based on an approximate separability of the 
interacting $N$-fermion Schr\"odinger equation in the hyperradial coordinate, 
which sheds light on the emergence of Laughlin states as 
well as other FQH states. An approximate good collective quantum number, 
the grand angular momentum $K$ from $K$-harmonic few-body theory, is shown 
to correlate with known FQH states at many filling factors observed 
experimentally.  
%We propose that these states could be observed 
%experimentally by confining weakly repulsive 2D spin-polarized fermions, 
%e.g. in a harmonic trap or a tight optical lattice, even in the absence 
%of any magnetic field.
\end{abstract}


\maketitle

\section{Introduction}
One of the most striking aspects of nonrelativistic quantum mechanics in 
more than one dimension is the remarkable implication of high 
degeneracy or near-degeneracy.  Textbook examples include the 
sp-hybridization of chemical bonds and the degenerate Stark effect of 
excited hydrogen atoms.  In the degenerate Stark effect, for instance, 
even an infinitesimally small external electric field 
selects energy eigenstates that are linear combinations of a finite set 
of degenerate zero-field states, and these are the same eigenstates that 
can be obtained by separating the field-free Schr\"odinger equation for 
that system in parabolic coordinates. 

A major development presented in 
this article is that similar considerations apply to the fractional 
quantum Hall effect (FQHE)~\cite{Tsui1982,Stormer1999}. 
In the fractional quantum Hall effect,
a strongly-interacting two-dimensional electron gas
exhibits quantization in the presence of a strong, perpendicular magnetic field.
In the noninteracting limit, the electrons fall into highly degenerate Landau levels,
and their collective behavior depends on the filling factor,
the ratio of the number of electrons to the large, but finite, degeneracy of the 
lowest Landau level for a sample with finite area.
The system exhibits quantization when the filling factor takes on integer or certain rational fraction values.
We show that the high degeneracy of the noninteracting system
produces dramatic implications.

Moreover, we demonstrate that the N-electron 
Schr\"odinger equation is approximately separable in hyperspherical 
coordinates.  This approach to the problem shows that a characteristic 
property of the noninteracting system, which we denote the {\it 
exceptional degeneracy}, becomes unusually high for precisely those 
states that appear at experimentally and theoretically observed FQHE 
filling factors.  In other words, even though the FQHE is viewed 
fundamentally as the epitome of a strongly-correlated system of 
electrons, {\it the occurrence or nonoccurrence of a FQHE filling factor 
is highly correlated with the pattern of exceptional degeneracies in the 
noninteracting electron system}.  The approximate separability 
demonstrated in hyperspherical coordinates for the few-body quantum Hall 
states makes definite predictions about a class of excitation 
frequencies that could be used to experimentally probe the system.

Extensive progress in the theoretical understanding of the FQHE has been 
achieved through various approaches following the early intuitive 
development by Laughlin~\cite{Laughlin1983}, notably the composite fermion (CF) picture 
developed by Jain~\cite{Jain1989, jainbook}, and work by Haldane~\cite{Haldane1983}, 
and Halperin~\cite{Halperin1984}.  Theoretical treatments have tended 
to reside in one of two different categories, either postulating trial 
wavefunctions as in Laughlin's original approach, (e.g.~\cite{Moore1991, Ginocchio1996, Wojs2004}) 
or else performing large numerical diagonalizations for the maximum number of electrons that can still give 
a manageable size computation, typically 8-20 particles (e.g.~\cite{Haldane1985}).  
A more recently developed technique uses CF wave functions as a basis 
for numerical diagonalization to study systems with 
larger numbers of particles~\cite{Jeon2007,Mukherjee2014}.  

The approach developed here has some advantages complementary 
to previous methods. 
In contrast to techniques that use the single-particle representation 
(i.e. Slater determinant constructions),
the approach treated here inherently uses collective coordinates.
It also provides a systematic expansion that can in principle describe 
any states existing in the Hilbert space of a finite number of 
particles, while at the same time allowing us to see many key properties 
analytically or with small-scale diagonalizations.  The adiabatic 
hyperspherical representation capitalizes on an approximate 
separability, and its key element is a set of potential energy curves 
showing at a glance the relevant size and energy of different energy 
eigenstates.  While the hyperradial degree of freedom is not separable 
for arbitrarily strong Coulomb interactions, our calculations 
demonstrate that approximate Born-Oppenheimer separability is an 
excellent approximation in typical regimes of electron density and field 
strength for a typical material like GaAs.

While the adiabatic hyperspherical 
representation~\cite{Macek1968JPB,fano1981,fano1983} has not been used extensively in 
condensed matter theory, it has had extensive success in a wide range of 
few-body contexts.  The literature in this field documents theoretical 
results that have been achieved in contexts as diverse as nuclear 
structure and 
reactivity~\cite{smirnov1977Sov.J.Part.Nucl.,avery1989, avery1993JPC,nielsen2001PRep,DailyKievskyGreene2015}, 
universal Efimov physics in cold atoms and 
molecules~\cite{WangDIncaoEsry2013,Rittenhouse2011JPB, Wang2014arXiv1412p8094,efimov1970PLB, nielsen1999PRLb,gattobigio2012PRA}, few-electron 
atoms~\cite{lin1986AMOP,lin1995PRep, lin2000PHYSICSESSAYS}, and systems containing positrons 
and electrons~\cite{botero1985PRA,DailyGreene2014,DailyGreene2015PRA,archer1990PRA}. Efimov's prediction~\cite{efimov1970PLB} of a universal binding mechanism for three particles at very large scattering lengths can itself be viewed as an application of the adiabatic hyperspherical coordinate treatment in a problem where the method is exact, although Efimov did not himself express it in those terms. Hyperspherical coordinates have also been 
employed to describe some many-body phenomena such as 
the trapped atom Bose-Einstein condensate with either attractive or 
repulsive interactions~\cite{BohnEsryGreene1998PRA,kushibe2004PRA} and the trapped 
degenerate Fermi gas in three dimensions including the BCS-BEC crossover 
problem.~\cite{rittenhouse2006PRA,Rittenhouse2011JPB} 

Our initial presentation of the formulation begins by setting up the 
problem rigorously for N electrons confined to a plane with a transverse 
uniform magnetic field. The antisymmetric states for spin-polarized fermions
are found directly using the technique developed in Ref.~\cite{efros1995}. 
Next we show that the exact separability of the Schr\"odinger equation 
in hyperspherical coordinates for the case of noninteracting electrons 
still exhibits an approximate separability even in the presence of 
Coulomb interactions.  The treatment then demonstrates how interactions single out 
potential energy curves or channels of exceptional degeneracy, which 
correlate with known Laughlin states, composite fermion FQHE states, and 
suggest other states that deserve future theoretical and experimental 
investigation.  This enables further predictions of a class of 
excitation frequencies that should be experimentally observable in a 
FQHE experiment.

This paper is organized as follows.  Sections \ref{sectionII} and \ref{sectionIII} formulate 
the one-body and N-body relative Hamiltonians in the symmetric gauge. Section \ref{sectionIV}
defines the hyperspherical coordinates adopted in the present study and 
writes the unsymmetrized hyperspherical harmonics which serve as our
primitive basis set.  This basis set is then connected to the Landau 
level picture and suggests a definition for the hyperspherical filling 
factor. The effect of Coulomb interactions is then developed and 
treated within the adiabatic hyperspherical representation.  Section \ref{sectionV} 
introduces the concept of exceptional degeneracy and computes this key 
quantity, which correlates with filling factors that are observable as 
FQH ground states.  Section \ref{sectionVI} offers concluding remarks and comments on 
future directions.  Finally, Appendix~\ref{appendix} relates the states with different 
hyperspherical filling factors to several of the states that have
previously been identified in the conventional composite fermion picture.

\section{Single particle Hamiltonian,\label{sectionII}}
The noninteracting Hamiltonian $H$ for a single electron
in an external magnetic field is given by
\begin{align}
\label{eq_H1}
H = \frac{1}{2m_e} \left( -\imath \hbar \bm{\nabla} 
                           + e \bm{A}\right)^2
\end{align}
in SI units where $m_e$ is the effective mass of the electron in the medium,
$e$ is the magnitude of the electron charge,
$\hbar$ is Planck's constant,
and $\bm{A}$ is the vector potential. 
For two dimensional space,
in Cartesian coordinates,
the gradient is 
$\bm{\nabla}=\hat{x} \partial_{x}  + \hat{y} \partial_{y} $.
For a constant magnetic field of magnitude $B$
oriented in the positive $\hat{z}$ direction,
the vector potential is $\bm{A}= (B/2)(-y \hat{x}+ x \hat{y})$.
Expanding Eq.~\eqref{eq_H1} with this choice of $\bm{A}$ yields
\begin{align}
  \label{eq_Hsingle}
  H = - \frac{\hbar^2}{2m_e} \bm{\nabla}^2 
  + \frac{e^2 B^2}{8m_e }(x^2+y^2)
  + \frac{e B}{2m_e} L_{z}, 
\end{align}
where $L_{z}$ is the $z$-component of the angular momentum operator,
$L_{z} = -\imath\hbar(x \partial_{y} - y \partial_{x})$.

The rest of this paper uses magnetic units where length 
is expressed in units of $\lambda_0$,
\begin{align}
  \label{eq_maglength}
  \lambda_0 = \sqrt{\frac{\hbar}{m_e \omega_c}}
\end{align}
and energy is expressed in units of $\hbar \omega_c$, where
$\omega_c$ is the cyclotron frequency, $\omega_c=eB/m_e$.
In these units, 
expressing $H$ in polar coordinates yields
\begin{align}
  \label{eq_H1r}
  H = -\frac{1}{2} \left\{ \frac{1}{r} \partial_r %\frac{d}{dr}
  r \partial_r %\frac{d}{dr}
  - \frac{L_z^2}{\hbar^2 r^2}\right\}
  + \frac{1}{8} r^2
  + \frac{1}{2\hbar} L_{z},
\end{align}
and the single-particle energy $E^{(1)}$ is
\begin{align}
  \label{eq_NIenergy}
  E^{(1)} = \frac{1}{2} \left(2 n + m + |m| + 1 \right),
\end{align}
where $n$ is a nodal quantum number
and $m$ is the rotational quantum number about the $z$-axis.
Section~\ref{sec_HN}
shows how the Hamiltonian is modified
when including more degrees of freedom,
while Sec.~\ref{sec_HNI}
makes additional modifications 
when expressed in hyperspherical coordinates.
Many aspects of Eqs.~\eqref{eq_H1r} and~\eqref{eq_NIenergy}
carry over to this formalism.

\section{N-body relative Hamiltonian\label{sectionIII}}
\label{sec_HN}
The $N$-body noninteracting
Hamiltonian $H_N$ is separable into a center of mass
$H_{\rm CM}$ and relative $H_{\rm rel}$ components,
\begin{align}
  H_N = H_{\rm CM} + H_{\rm rel},
\end{align}
where, in Cartesian coordinates akin to Eq.~\eqref{eq_Hsingle},
\begin{align}
  \label{eq_HN}
  H_{\rm rel} = &- \frac{1}{2\mu}\sum_{j=1}^{N_{\rm rel}} \bm{\nabla}_j^2 
  + \frac{\mu}{8}\sum_{j=1}^{N_{\rm rel}} (x_j^2 + y_j^2) 
  + \frac{1}{2\hbar}\sum_{j=1}^{N_{\rm rel}} L_{z_j}^{\rm rel}.
\end{align}
Here,
$N_{\rm rel}=N-1$ is the number of relative Jacobi vectors $\bm{\rho}_j$
with Cartesian components $x_j$ and $y_j$,
and $\mu$ is a dimensionless mass scaling factor~\cite{delves1959NP,delves1960NP},
\begin{align}
  \label{eq_mu}
  \mu = \left( \frac{1}{N} \right)^{1/N_{\rm rel}}.
\end{align}
The center-of-mass Hamiltonian is similar in form to Eq.~\eqref{eq_HN},
except $\mu$ is replaced by $N$ and %,
there is only the center-of-mass vector $\bm{\rho}_{\rm CM}$.

\begin{figure}
  \centering
  \includegraphics[angle=0,width=0.4\textwidth]{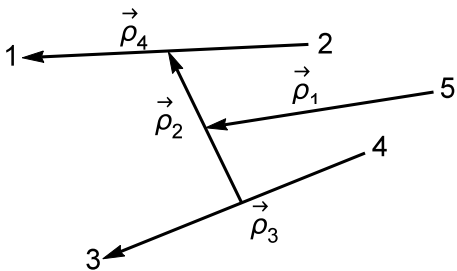} 
  \caption{(Color online) 
    Diagram describing the mass-scaled Jacobi coordinates
    of Eq.~\eqref{eq_jmatrix}.
  }
  \label{fig_Jtree}
\end{figure}

The linear transformation from single-particle to center-of-mass
and relative Jacobi vectors is arbitrary,
but in this work,
the scheme used first joins identical particles into pairs,
then joins the center of mass of each pair into ever larger clusters.
For odd $N$, the unpaired electron is joined to 
the center of mass of the other paired particles.
The Jacobi vectors are labeled in a reverse manner,
so that the last Jacobi vector, the $(N-1)^{st}$, is
always the relative coordinate for a pair of particles,
and where the Jacobi vectors of the largest clusters have 
the smallest index.
For example,
for five electrons
in matrix notation this is
\begin{align}
  \label{eq_jmatrix}
  \begin{pmatrix} 
    \bm{\rho}_1 \\ \bm{\rho}_2 \\ \bm{\rho}_3 \\ \bm{\rho}_4 \\ \bm{\rho}_{\rm CM}
  \end{pmatrix} = 
  \left( \!\!\!\!
  \begin{array}{rlcccr}
    \sqrt{\frac{4/5}{\mu}}  \times
    &\{ \frac{1}{4}
    & \frac{1}{4} 
    & \frac{1}{4}
    & \frac{1}{4}
    & -1\} \\
    \sqrt{\frac{1}{\mu}}  \times
    &\{ \frac{1}{2}
    & \frac{1}{2} 
    & -\frac{1}{2}
    & -\frac{1}{2}
    & 0\} \\
    \sqrt{\frac{1/2}{\mu}}  \times
    &\{ 0
    & 0
    & 1
    & -1
    & 0\} \\
    \sqrt{\frac{1/2}{\mu}}  \times
    & \{ 1
    & -1
    & 0
    & 0
    & 0\} \\    
    \frac{1}{5} \times & \{ 1 & 1 & 1 & 1 & 1\}
  \end{array} \right) \!\!
  \begin{pmatrix}
    \bm{r}_1 \\ \bm{r}_2 \\ \bm{r}_3 \\ \bm{r}_4 \\ \bm{r}_5
  \end{pmatrix} \!\! .
\end{align}
Figure~\ref{fig_Jtree}
is a diagrammatic representation of the Jacobi vectors described
by Eq.~\eqref{eq_jmatrix}.
The numbers denote particle locations while
the arrows denote the Jacobi vectors,
also labeled by $\bm{\rho}_j$.
This choice of Jacobi tree
reduces the size of the unsymmetrized basis needed
to achieve antisymmetric states,
as is described in Sec.~\ref{sec_relativeH}.

\section{Hyperspherical form\label{sectionIV}}
This section describes the hyperspherical transformation
of the noninteracting relative Hamiltonian,
how the relative Hamiltonian is expressed in these coordinates,
and the resulting adiabatic potentials.

\subsection{Hyperspherical coordinate transformation}
\label{sec_HSCT}
Hyperspherical coordinates are the generalization of
spherical coordinates beyond three degrees of freedom.
The size of the system is correlated with a single length,
the hyperradius $R$,
while the geometry of the system is encoded
in the remaining degrees of freedom as a set of hyperangles,
denoted by $\bm{\Omega}$.
This length $R$,
\begin{align}
  \label{eq_hyperradius}
  R^2=\sum_{j=1}^{N_{\rm rel}} \rho^2_j,
\end{align}
is a scalar quantity
and its square is the sum of the squared lengths of the Jacobi vectors.

\begin{figure}
  \centering
  \includegraphics[angle=0,width=0.4\textwidth]{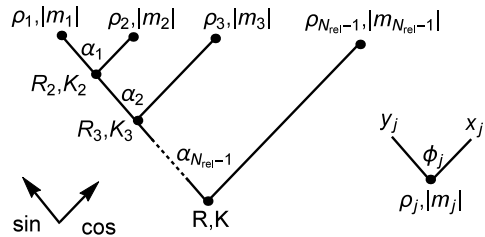} 
  \caption{(Color online) 
    Semi-canonical Jacobi tree diagram.
    The sub-tree in the lower right describes the $N_{\rm rel}-1$
    nodes at the top of the main tree.
    Reading from the main node at $R,K$,
    any time a node is passed to the left (right),
    the coordinate picks up a factor of $\sin\alpha_j$ ($\cos\alpha_j$).
  }
  \label{fig_Jtree2}
\end{figure}

The orthogonal coordinate transformation from Cartesian to hyperspherical 
coordinates has some arbitrariness,
and many different schemes can be found
in the literature~\cite{smirnov1977Sov.J.Part.Nucl.,avery1989,avery1993JPC,Rittenhouse2011JPB}.
In this work,
the semi-canonical construction is most useful [see also Fig. A4 of Ref.~\cite{Rittenhouse2011JPB}].
It resembles the canonical tree structure of Refs.~\cite{smirnov1977Sov.J.Part.Nucl.,avery1989},
but instead of each new branch adding a single degree of freedom,
each additional branch adds an additional two-dimensional sub-tree.
In this way,
the particle-like nature of the two-dimensional Jacobi vectors
is maintained.
This set of hyperangles consists of
the $N_{\rm rel}$ azimuthal angles $\phi_j$
associated to each Jacobi vector $\bm{\rho}_j$
and the $N_{\rm rel}-1$ constructed hyperangles $\alpha_j$,
where
\begin{align}
  \label{eq_alpha}
  \tan \alpha_j = \frac{\sqrt{\sum_{k=1}^j \rho_k^2}}{\rho_{j+1}}.
\end{align}
Figure~\ref{fig_Jtree2}
gives a diagrammatic representation of the semi-canonical construction.
This Jacobi tree connects branches (segments) into nodes (dots),
where to every node is associated a sub-length,
a sub-hyperangular quantum number [see e.g. Eq.~\eqref{eq_subh}],
and an angle.
%% \begin{align}
%%   \tan \alpha_j = \frac{\sqrt{\sum_{k=1}^j \rho_k^2}}{\rho_{j+1}}.
%% \end{align}
The sub-lengths $R_j$ are defined similarly to Eq.~\eqref{eq_hyperradius},
e.g. $R_2^2 = \rho_1^2 + \rho_2^2$. 

The Jacobi tree contains all of the information describing
the coordinate transformation from Cartesian to hyperspherical.
Read from the main node at $\{R,K\}$ to the top nodes at $\{\rho_j,|m_j|\}$,
any time a node is passed to the left (right),
the Cartesian coordinate picks up a factor of $\sin\alpha_j$ ($\cos\alpha_j$).
The sub-tree in the lower right of Fig.~\ref{fig_Jtree2}
describes the $N_{\rm rel}-1$ nodes at the top of the main tree.
For example,
$x_1 = \rho_1 \cos \phi_1$.
However,
for $N=5$,
$x_1$ in terms of every node is expressed as
$x_1 = R \sin\alpha_3 \sin \alpha_2 \sin \alpha_1 \cos \phi_1$.
The $\phi_j$ angles range from $0<\phi<2\pi$
while the hyperangles $\alpha_j$ range from $0<\alpha_j<\pi/2$.
The volume element for each $\phi_j$ is $d\phi_j$,
while for each $\alpha_j$ it is $\sin^{2j-1}\alpha_j \cos\alpha_j d\alpha_j$.

\subsection{Relative Hamiltonian}
\label{sec_relativeH}
The noninteracting relative Hamiltonian,
Eq.~\eqref{eq_HN},
transforms to 
\begin{align}
  \label{eq_HNhs}
  H_{\rm rel} = - \frac{1}{2\mu} \bm{\nabla}^2_{R, {\bm \Omega}} 
  + \frac{\mu}{8} R^2 
  + \frac{1}{2\hbar}L_z^{\rm rel,tot}
\end{align}
where $L_z^{\rm rel,tot}$ is the total relative $z$-component
of the angular momentum,
and the Laplacian operator in hyperspherical coordinates
$\bm{\nabla}^2_{R, {\bm \Omega}}$ becomes
\begin{align}
  \bm{\nabla}^2_{R, {\bm \Omega}} =   \frac{1}{R^{2N_{\rm rel}-1}} \partial_R
    R^{2N_{\rm rel}-1} \partial_R
    - \frac{\hat{\bm{K}}^2}{R^2}.
\end{align}
$\hat{\bm{K}}$ is called the grand angular momentum operator~\cite{avery1989}.
Note the similarity with Eq.~\eqref{eq_H1r},
where there are radial and angular components.

The eigenstates of $\hat{\bm{K}}^2$ are the hyperspherical harmonics,
$\Phi_{K{u}}^{(M)}(\bm{\Omega})$, where
\begin{align}
\hat{\bm{K}}^2 \Phi_{K{u}}^{(M)}(\bm{\Omega}) 
   = K (K + 2N_{\rm rel} - 2)\Phi_{K{u}}^{(M)}(\bm{\Omega})
\end{align}
and the set of hyperspherical harmonics are orthonormal 
over the hyperangles $\bm{\Omega}$,
\begin{align}
\int d\bm{\Omega}\; \Phi_{K'{u}'}^{(M)*}(\bm{\Omega})
                  \Phi_{K{u}}^{(M)}(\bm{\Omega}) 
   = \delta_{K'K}\delta_{{u}'{u}}.
\end{align}
The grand angular momentum quantum number $K$ ($K=0,1,2,\ldots$)
is analogous to the angular momentum quantum number;
for the noninteracting system, it remains a good quantum number.
Here, $u$ is an index used to label the different unsymmetrized states within
a given $K$-manifold (fixed $K$ subspace).
For $2N_{\rm rel}$ degrees of freedom, 
there are $(2N_{\rm rel}-2+2K)(2N_{\rm rel}-3+K)!/(K !(2N_{\rm rel}-2)!)$
linearly independent unsymmetrized functions.

The projection quantum numbers $m_j$ associated with the Jacobi vectors
are not good quantum numbers
in the presence of interactions and antisymmetrization,
however the total projection quantum number $M$ remains a good quantum number
of the system,
even with Coulomb interactions (or any other interactions that
depend only on the inter-particle distances).
The set of hyperspherical harmonics are made 
simultaneous eigenstates of  $\hat{\bm{K}}^2$ and $L_z^{\rm rel,tot}$ 
by enforcing the constraint
\begin{align}
  \sum_{j=1}^{N_{\rm rel}} m_j = M,
\end{align}
$|M| \le K$,
which is assumed in the following discussion.
Because the center of mass has already been separated,
it does not contribute to $M$, and its $L_z$ and energy can be incorporated into 
the system trivially.

In the semi-canonical coupling scheme of this work (see Fig.~\ref{fig_Jtree2}),
the unsymmetrized hyperspherical harmonics are expressed as
%TRIPLE CHECK THESE
\begin{widetext}
  \begin{align}
    \label{eq_HH}
    \Phi_{K{u}}^{(M)}(\bm{\Omega}) = &
    \prod_{j=1}^{N_{\rm rel}} \frac{e^{\imath m_j \phi_j}}{\sqrt{2\pi}}
    \prod_{k=1}^{N_{\rm rel}-1} {\cal N}_k \; \sin^{K_k}\alpha_k \;%\times 
    %\\ \nonumber &
    \cos^{|m_{k+1}|}\alpha_k
    \; P_{n_k}^{K_k+(k-1),|m_{k+1}|}(\cos 2\alpha_k ),
  \end{align}
  \begin{align}
    {\cal N}_k^2 = \frac{(2 K_{k+1} + 2k)
      \; \Gamma(n_{k}+K_{k}+|m_{k+1}|+k)
      %\Gamma([K_k-K_{k-1}-|m_k|]/2+1)}
      \; \Gamma(n_{k}+1)}
    %{\Gamma([K_k+K_{k-1}-|m_k|+2k]/2) \Gamma([K_k+|m_k|-K_{k-1}+2]/2)}, 
    {\Gamma(n_{k}+K_{k}+k) \; \Gamma(n_{k}+|m_{k+1}|+1)}, 
  \end{align}
\end{widetext}
where the $P$ are Jacobi polynomials,
${\cal N}_k$ is the normalization for each Jacobi polynomial,
and the $K_k$ are sub-hyperangular ``quantum numbers,''
defined recursively as
\begin{align}
  \label{eq_subh}
  K_1 & = |m_1|, \mbox{ and}\\
  K_k & = 2 n_{k-1} + K_{k-1} + |m_k|.
\end{align}
The $n_k$ ($n_k=0,1,2,\ldots$) are determined after fixing the various $K_k$
and $|m_k|$.
In practice,
it is easier to first choose the $n_k$ and $|m_k|$,
since $n_k$ is the order of the Jacobi polynomial,
then determine the $K_k$.

With the exception of the grand angular momentum $K$ and the 
total azimuthal quantum number $M$, 
neither the $K_k$ nor the $m_k$ remain good quantum numbers after antisymmetrizing
the hyperspherical harmonics.
We antisymmetrize the set of functions Eq.~\eqref{eq_HH}
following the method of Ref.~\cite{efros1995}.
In a closed subspace of the Hamiltonian
(here, the subspaces with both fixed $K$ and fixed $M$),
the antisymmetrized states must be linear combinations
of the unsymmetrized basis functions within the same subspace. 
As before, the subscript ${u}$ is an index that
distinguishes different orthogonal basis functions in the same manifold;  
each index ${u}$ is associated with a different set of good hyperspherical quantum numbers
that satisfy Eqs.~\eqref{eq_subh}.
Similarly, the new subscript $a$ will later be used 
to index different antisymmetric states in the same manifold, 
but in this case, the $K_k$ and $m_k$ no longer constitute good quantum numbers. 
The complete set of basis function labels in the unsymmetrized basis 
will be indicated by a bold $\bm{u} = \{1, 2, \ldots \}$,
and the number of basis functions will be given by $|\bm{u}|$.

The antisymmetric basis functions can be constructed by first
building the full matrix of the antisymmetrization operator 
$\hat{\mathcal{A}} =1-\hat{P}_{12}-\hat{P}_{13}-\hat{P}_{23}+\hat{P}_{123}+...$
(all $N!$ terms)  connecting all
unsymmetrized states $\Phi _{Ku}^{(M)}$ in a given $K,|M|$ manifold. 
This Hermitian matrix 
$\mathcal{A}_{ij}=\left\langle \Phi _{Ki}^{(M)}|\mathcal{A} |\Phi _{Kj}^{(M)}\right\rangle $ 
is then diagonalized, i.e. we find the eigenvalues
and their corresponding eigenvectors, $X_{ia}$. 
The $\overline{\mathcal{N}_a}$ eigenvectors with eigenvalues equal to $N!$ 
give the totally antisymmetric hyperangular wave functions,
\begin{align}
\Phi _{Ka}^{(M)}=\sum_{u}\Phi _{Ku}^{(M)}X_{ua}, \quad \quad a=1,2,...
\overline{\mathcal{N}_{a}},
\end{align}
where $\overline{\mathcal{N}_{a}}$ is the number of antisymmetric states and is typically smaller 
than the total dimension of the degenerate unsymmetrized subspace.
Note that the matrix $X$ depends of course on $K$ and $M$, 
but this has been suppressed for notational brevity.  

The most time-consuming part of this calculation is the determination of the
matrix $\mathcal{A}_{ij}$, which can be accomplished using a technique
proposed by Efros~\cite{efros1995}.
Instead of being calculated directly, the antisymmetrization matrix 
can be found by treating $\mathcal{A}_{ij}$ as many unknowns
in a linear system of equations.
The antisymmetrization matrix is first re-expressed in terms of a matrix equation,
\begin{align}
\label{eq_antimat}
{\cal \hat{A}}\Phi_{K j}^{(M)}(\bm{\Omega})  
=\sum_{i } \Phi_{K i}^{(M)}(\bm{\Omega}) \mathcal{A}_{ij},
\end{align}
where $i,j \in {\bm u}$ and $\Phi_{K j}^{(M)}(\bm{\Omega})$ represents a $|\bm{u}|$-length row array
of all of the unsymmetrized basis functions. 
The matrix $\mathcal{A}_{ij}$ is the unknown $|\bm{u}| \times |\bm{u}|$ matrix representation
of the antisymmetrization operator, and Eq.~\eqref{eq_antimat} 
evaluated at a single set of $N$-particle coordinates constitutes a system of $|\bf{u}|$ 
equations with ${\bm u^2}$ unknowns.
The antisymmetrization operator is understood to act on the particle coordinates, so that 
${\cal \hat{A}}\Phi_{K j}^{(M)}(\bm{\Omega})  = \Phi_{K j}^{(M)}({\cal \hat{A}}\bm{ \Omega}) $

Because Eq.~\eqref{eq_antimat} holds true for any set of N-particle coordinates, 
substituting any random set of angular coordinates ${\bm \Omega}_\gamma$ 
produces a different, linearly independent equation in the unknowns of the
antisymmetrization matrix, $\mathcal{A}_{ij}$.
Substituting $|\bf{u}|$ different sets of $N$-particle coordinates into
Eq.~\eqref{eq_antimat} results in a linear system of equations that can be solved 
for $\mathcal{A}_{ij}$.
If we write $\Phi_{\gamma j}$ to mean the two-dimensional array of unsymmetrized basis
functions evaluated at different sets of $N$-particle coordinates,
where the column index $j$ indexes the $|{\bm u}|$ different unsymmetrized basis functions, 
the row index $\gamma$ indexes the $|{\bm u}|$ different sets of coordinates, ${\bm \Omega}_\gamma$,
and the $K$ and $M$ quantum numbers have been suppressed, 
then $\mathcal{A}_{i j}$ can be found by solving the following equation:
\begin{align}
\label{eq_linsys}
\hat{\mathcal{A}} \Phi_{\gamma j} = \sum_i \Phi_{\gamma i} \mathcal{A}_{i j}.
\end{align}

In practice, constructing the matrices $\Phi_{\gamma i}$ and 
$\hat{\mathcal{A}} \Phi_{\gamma j}$ is trivially parallelizable,
but the memory requirements are significant 
since it requires the storage
of two $|\bm{u}| \times |\bm{u}|$ double complex dense matrices.
The size of the unsymmetrized basis increases dramatically 
with the number of particles.
Choosing paired Jacobi coordinates helps limit the growth 
in spin-polarized fermion systems,
since states with pair coordinates associated with even $m_j$ 
can be eliminated due to symmetry. 
However, the growth is still rapid,
and as a result we have only performed calculations in systems
with up to 6 spin-polarized electrons at $1/3$ filling.
In the future,
with programs like SCALAPACK,
it is feasible to push this analysis further.
The problem of antisymmetrizing the hyperspherical functions
becomes very challenging as unsymmetrized basis expands, 
and other strategies for antisymmetrization may also prove
effective~\cite{Barnea1999,Krivec1998, Viviani1998}.

\subsection{Eigenstates of the noninteracting $H_{\rm rel}$}
\label{sec_HNI}
Each exact eigenfunction $\Psi(R,\bm{\Omega})$
of the relative Hamiltonian, Eq.~\eqref{eq_HNhs},
is separable into a hyperradial function, $F_{n_RK}^{(M)}(R)$,
and one antisymmetrized hyperspherical harmonic ,$\Phi_{Ka}^{(M)}(\bm{\Omega})$,
\begin{align}
  \label{eq_wf}
  \Psi(R,\bm{\Omega}) = R^{-N_{\rm rel}+1/2}
   F_{n_RK}^{(M)}(R)\Phi_{K{a}}^{(M)}(\bm{\Omega}).
\end{align}
The many-dimensional hyperradial Schr\"odinger equation
thus reduces to a one-dimensional
uncoupled ordinary differential equation:
\begin{align}
  \label{eq_HNI}
  & \left\{ -\frac{1}{2\mu} \frac{d^2}{dR^2}
  + U_{K}^{(M)}(R) - E \right\} F_{n_RK}^{(M)}(R)
  %% \\ \nonumber 
  %% & + \sum_{K'q'}\frac{\langle K'q'|
  %%   C(\bm{\Omega}) |Kq\rangle}{R}F_{n_RK}^{(M)}(R)
  = 0,
\end{align}
where the noninteracting potentials $U_{K}^{(M)}(R)$
are given by
\begin{align}
  \label{eq_UNI}
  U_{K}^{(M)}&(R) =  \\ \nonumber
  &\frac{(K+N_{\rm rel}-1/2)(K+N_{\rm rel}-3/2)}{2\mu R^2}
  %% \bigg\{ K (K + 2N_{\rm rel} - 2) \\ \nonumber
  %% & + \frac{(2N_{\rm rel}+1)(2N_{\rm rel}+3)}{4} \bigg\}
  + \frac{\mu}{8}R^2 + \frac{1}{2}M.
\end{align}

The noninteracting hyperradial solutions $F_{n_RK}^{(M)}(R)$
to the scaled Schr\"odinger equation, Eq.~\eqref{eq_HNI}, are
\begin{align}
  F_{n_RK}^{(M)}(R) = {\cal N} %\exp\left(-\frac{\mu R^2}{4}\right)
  e^{-\tfrac{\mu R^2}{4}}
  L_{n_R}^{K+N_{\rm rel}-1}\left(\frac{\mu R^2}{2}\right) 
  R^{K+N_{\rm rel}-1/2},  \label{eq.FnRKMni}
\end{align}
where $n_R$ $(n_R=0,1,2,\ldots)$ is the hyperradial quantum number
and $L$ is the associate Laguerre polynomial.
The normalization, where $\int_0^{\infty} |F_{n_RK}^{(M)}(R)|^2dR=1$, is
\begin{align}
  {\cal N} = \sqrt{\frac{n_R! \; \mu^{K+N_{\rm rel}}}{\Gamma(n_R+K+N_{\rm rel}) \; 2^{K+N_{\rm rel}-1}}}.
\end{align}
The noninteracting many-body energies $E^{\rm NI}$ are
\begin{align}
  E^{\rm NI} = \frac{1}{2}(2n_R + M + K + N_{\rm rel}).
\end{align}
Note that this equation is similar in form
to Eq.~\eqref{eq_NIenergy}
where $n_R$ is a nodal quantum number
and $K$ plays the role of the $|M|$ term.
In the limit that $N_{\rm rel}=1$,
this equation reduces to Eq.~\eqref{eq_NIenergy}.

\begin{figure}
  \centering
  \includegraphics[angle=0,width=0.45\textwidth]{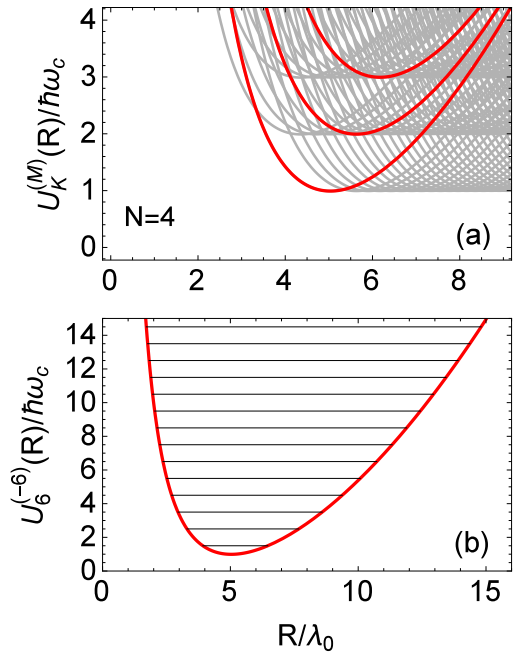} 
  \caption{(Color online) 
   (a) Light solid lines show the adiabatic potential curves
    of the noninteracting four-body system
    whose minima lie below $4\hbar\omega_c$.
    The lowest group ($K=|M|$) represents the lowest Landau level.
    The next and next-next higher groups are for
    $K=|M|+2$ and $K=|M|+4$, respectively.
    The dark solid lines are for $M=-6$,
    the lowest of which supports the integer quantum Hall state.
    (b) The $M=-6$ hyperradial noninteracting potential curve 
    in the lowest Landau level and the corresponding hyperradial energy spectrum.
  }
  \label{fig_pot}
\end{figure}

Figure~\ref{fig_pot}
shows the effective potentials $U_K^{(M)}(R)$, Eq.~\eqref{eq_UNI},
for the noninteracting system with four identical fermions. 
The light solid lines show all the potentials visible
in the given scale,
while the dark solid lines highlight those curves with
$M=-6$ and,
from bottom to top, $K=6, 8,$ and $10$, respectively.
One striking feature is that the potentials group into ``bands''
whose potential minima group around the same energy.
Each band is separated by a cyclotron unit of energy,
reminiscent of Landau levels.
However,
these potentials are not single-particle potentials, 
but rather support many-body states.
In the case of Figure~\ref{fig_pot}a,
energy arguments indicate the first excited ``band''
represents a four-body state with a filled lowest Landau level
and an excited electron.

For any $K,M$,
the particles are confined due to the diamagnetic term and
each of these potentials supports an infinite number of bound states.
The noninteracting hyperradial solutions, Eq.~\eqref{eq.FnRKMni},
are centered within the non-inteacting potential wells, $U_K^{(M)}(R)$.
For a given potential,
each nodal excitation in the hyperradius adds one unit of cyclotron energy
$\hbar \omega_c$, as shown in Fig.~\ref{fig_pot}b,
and the additional hyperradial nodes increase the overall size of the hyperradial wave function
due to the additional hyperradial nodes.
For a fixed $M$,
the allowed values of $K$ are $K=|M|,|M|+2,\ldots$
in step sizes of 2.
For each increase in $K$,
the confining potential moves out in the hyperradius,
indicating the size of the system is increasing
as the energy increases.

Another striking feature is the potential
that corresponds to the integer quantum Hall state.
The dark curve with $K=|M|$,
whose minimum is about $R\approx5\lambda_0$,
is isolated from the other curves with higher $K=|M|$ values.
In fact, there is always a single isolated curve separated from the
other curves at $K = |M| = N(N-1)/2$ for all system sizes.
The one antisymmetric hyperspherical wave function of this manifold,
when re-expressed in terms of independent particle coordinates, 
is exactly identical to the wave function of the integer quantum Hall state.
Moreover,
the first excited ``band'',
representing a filled lowest Landau level
and an excited electron,
allows potentials at smaller $K$ values.
Exciting a single electron to a higher Landau level, 
allows the system to compress to a smaller hyperradius 
at the cost of one $\hbar \omega_c$ unit of energy.

\subsection{Hyperspherical filling factor}
The most important parameter in the quantum Hall problem is the filling factor, $\nu$, the number of occupied Landau levels in a sample in the noninteracting limit. It is given by 
\begin{align}
\nu = \frac{\rho h}{ e B} = \frac{N \phi_0}{B A}, \label{eq_fillingfraction} 
\end{align}
where $\rho$ is the two-dimensional electron density, $B$ is the magnetic field, $N$ is the number of electrons, $A$ is the sample area, and $\phi_0$ is the fundamental flux quantum, $\phi_0 = h/e$ in S.I. units~\cite{jainbook}. The filling factor gives the ratio of the number of electrons in the system to the number of available single particle orbitals in the lowest Landau level for a given sample area. The electron density can be somewhat controlled with doping and gate voltages, but for typical experiments in gallium arsenide~\cite{Tsui1982,Pan2008} is on the order of $1-3\times 10^{11} cm^{-2}$. As an example, in a system with $\rho = 2.4\times 10^{11} cm^{-2}$~\cite{Eisenstein1990}, the $\nu = 1$ quantum Hall state is found at a magnetic field near $B = 10$T and the $\nu = 1/3$ state occurs around the much higher field $B \approx 29$T.

Experimentally, the Hall resistance quantizes to values of $R_H = h/\nu e^2$. Defining the filling factor in terms of the electron density is ideal for experimental systems, where the local electron density is averaged over enormous numbers of electrons. However, for systems with few electrons, there is some reasonable ambiguity in establishing the average density of a sample that is not sharply confined. In most numerical models of the planar system, the ambiguity about the area is resolved by cutting out of the Hilbert space all single-particle wave-functions whose maxima in the lowest Landau level lies outside of a certain radius. The area of the disk defined by this radius is an approximate area for the few-particle model system. 

In the hyperspherical construction, the area of a distribution of $N$ identical mass particles correlates with the particles' hyperradius, Eq.~\eqref{eq_hyperradius}, which can be used to define the filling factor without reference to the single particle wave functions. The connection between the hyperradius and the area can be established by a statistical average over a large ensemble of systems, holding the particle number $N$, and the sample radius, $r_c$ constant. Each individual system in the ensemble consists of a random distribution of $N$ particles over a disk with radius $r_c$. The particle density over this area is clearly $\rho = N/\pi r_c^2$, although the density is obviously non-uniform.  

In general, systems with different random distributions of $N$ particles will have different hyperradii, but the value of the hyperradius squared tends to increase on average with increasing particle density and tends to decrease with disk size. The statistical average of the hyperradius squared for a fixed $N$ and $\pi r_c^2$ over many trials is empirically found to be
\begin{align}
\langle R^2 \rangle_{N, r_c}= \frac{(N-1) r_c^2}{2 \mu}, \label{eq_avgHRsq}
\end{align}
where $\mu$ is the reduced mass of $N$ particles, Eq.~\eqref{eq_mu}.  For any individual distribution of $N$ particles on a circle of radius $r_c$, Eq. \eqref{eq_avgHRsq} typically does not reproduce $r_c$ accurately. However, the definition of the particle density is also only described as an average. The particle density $\rho = N/\pi r_c^2$ for a few particles on a disk also does not actually give the average density of those particles except with reference to a statistical average. The hyperradius gives a consistent way to measure an approximate area of a distribution of particles, even in the less clearly defined case of very few particles.

Using Eq.~\eqref{eq_avgHRsq}, it is straightforward to find a hyperspherical expression for the filling factor. We first scale all lengths in the problem by the magnetic length of the system, $\lambda_0$ [see Eq.~\eqref{eq_maglength}], and get the filling factor in terms of the radius of a disk, $r_c$, that gives an approximate area of the sample.
Substituting for this characteristic disk radius, equation Eq. \eqref{eq_fillingfraction} becomes
\begin{align}
  \nu = \frac{N(N-1)}{ \mu \langle R^2 \rangle }. \label{eq_almostnu}
\end{align}

This expression for the filling factor can be used to distinguish the filling factors of different noninteracting wave functions.   In the lowest Landau level, the (unnormalized) noninteracting hyperradial wave functions have no hyperradial excitations and depend very simply on the hyperradius and the grand angular momentum:
\begin{align}
 R^{-N_{\rm rel} +1/2}  F_{0,K}^{(M)}(R)   \propto  e^{-\mu R^2/4} R^K.
\end{align}
The peak of this wave function is easily located at $R^2 = 2K/\mu$, and indicates a most-likely hyperradius for each noninteracting wave function.  Substituting this $R^2$ into Eq. \eqref{eq_almostnu} gives the following simple expression for the filling factor in the lowest Landau level,
\begin{align}
  \nu = \frac{N(N-1)}{2K}. \label{eq_fillingfactor}
\end{align}
This formula accurately determines the filling factor for the $\nu = 1$ and the Laughlin $\nu = 1/m$ states, for both boson and fermion systems. Other Jain composite fermion fillings are not accurately assigned by this function; instead the identified states occur at a small shift away from this ideal hyperspherical filling function (see Appendix~\ref{appendix}).  This shift is most prominent in few-body systems, and approaches zero in the thermodynamic limit.

\subsection{Coulomb interaction}
Including Coulomb interactions in
the noninteracting Hamiltonian Eq.~\eqref{eq_HNhs}
yields
\begin{align}
  \label{eq_HNhsC}
  H_{\rm rel} = - \frac{1}{2\mu} \bm{\nabla}^2_{\bm \Omega} 
  + \frac{\mu}{8} R^2 
  + \frac{1}{2\hbar}L_z^{\rm rel,tot}
  + \kappa %% \sqrt{\frac{1/2}{\mu}}
  \frac{C(\bm{\Omega})}{R},
\end{align}
where $\kappa$ is a dimensionless parameter
that determines the strength of the Coulomb interactions.  
Here,
\begin{align}
  \kappa = \frac{e^2}{4 \pi \epsilon \lambda_0}
  \frac{1}{\hbar\omega_c }
  = \frac{a_0}{\lambda_0}
    \frac{E_H}{\hbar\omega_c}
    \frac{1}{\epsilon/\epsilon_0}
  = \frac{1}{\epsilon/\epsilon_0} \sqrt{\frac{E_H}{\hbar\omega_c}},
\end{align}
where $a_0$ is the Bohr radius,
$E_H$ is the Hartree unit of energy, and
$a_0^2E_H=\lambda_0^2\hbar\omega_c=\hbar^2/(2 m_e)$
  with $a_0 E_H = e^2/(4\pi\epsilon_0)$.
$\epsilon$ is the permittivity of the material,
in units of the permittivity of free space $\epsilon_0$. In the lowest Landau level for typical experiments in gallium arsenide, $\kappa$ is on the order of one or smaller; for example, $\kappa \approx  0.76$ at $B = 10$T.

The form of the hyperspherical Coulomb term $C(\bm{\Omega})$
depends on the choice of Jacobi vectors and hyperangles.
In general,
from single-particle to relative coordinates
the Coulomb interaction involves a linear combination
of only the relative Jacobi vectors,
\begin{align}
  \sum_{i<j}\frac{\kappa}{|\bm{r}_i-\bm{r}_j|}
  \to \sum_{i<j}
  \frac{\kappa}
       {\sqrt{\mu} \; |\sum_{k=1}^{N_{\rm rel}} \beta_k^{ij} \bm{\rho}_k|}.
\end{align}
For a concrete example,
invert the transformation matrix Eq.~\eqref{eq_jmatrix}
to express the $\bm{r}_j$, $j=1\ldots 5$
in terms of the $\bm{\rho}_k$, $k=1\ldots 4$
and take vector differences.

Because we use the antisymmetrized hyperangular functions
the Coulomb interaction need only be calculated
between one pair of electrons,
then scaled by the number of pairs.
It is simplest to use the last Jacobi vector
that is proportional to the distance between a pair of electrons.
In the reverse construction of the hyperangles,
the last Jacobi vector is the simplest to express in hyperspherical
coordinates [see Eq.~\eqref{eq_alpha} and Fig.~\ref{fig_Jtree2}],
such that
\begin{align}
  \kappa \frac{C(\bm{\Omega})}{R}
  \to \frac{N (N-1)}{2} \sqrt{\frac{1}{2 \mu}}
  \frac{\kappa}{R \cos\alpha_{N_{\rm rel}-1}}.
\end{align}
This simplification is valid only when the basis functions are either totally symmetric or totally antisymmetric. 
Thus,
integrating the above expression 
in the basis of unsymmetrized hyperspherical harmonics
[see Eq.~\eqref{eq_HH}]
reduces to a one-dimensional integral in $d\alpha_{N_{\rm rel}-1}$
because the other integrations are accomplished
via orthogonality of the angular functions.
In practice,
Gauss-Jacobi quadrature is used to evaluate the
integral in $d\alpha_{N_{\rm rel}-1}$.

The strategy to diagonalize Eq.~\eqref{eq_HNhsC} remains the same
as that for the noninteracting system.
First,
$M$ remains a good quantum number
and each $M$ block of $H_{\rm rel}$ is diagonalized independently.
However,
the expansion Eq.~\eqref{eq_wf} is no longer strictly separable
into radial and hyperangular functions.
Instead,
the hyperangular channel functions depend parametrically on $R$,
where
\begin{align}
  \label{eq_wfc}
  \Psi(R,\bm{\Omega}) = R^{-N_{\rm rel}+1/2}
  \sum_{\chi} F_{E\chi}^{(M)}(R)\Phi_{\chi}^{(M)}(R;\bm{\Omega}).
\end{align}
Here,
$\chi$ labels each channel
and the channel functions $\Phi_{\chi}^{(M)}(R;\bm{\Omega})$
are orthonormal for a fixed hyperradius,
\begin{align}
\int d\bm{\Omega}\; \Phi_{\chi}^{(M)*}(R;\bm{\Omega})
                  \Phi_{\chi'}^{(M)}(R;\bm{\Omega}) 
   = \delta_{\chi\chi'}.
\end{align}

The hyperradius $R$ is treated as an adiabatic parameter,
where the adiabatic Hamiltonian $H_{\rm ad}$,
\begin{align}
  \label{eq_Had}
  H_{\rm ad} = & \frac{1}{2\mu R^2}
  \bigg\{ \hat{\bm{K}}^2 
  + (N_{\rm rel}-1/2)(N_{\rm rel}-3/2) \bigg\} \\ \nonumber
  %+ \frac{(2N_{\rm rel}-1)(2N_{\rm rel}-3)}{4} \bigg\} \\ \nonumber
  &+ \frac{\mu}{8}R^2 + \frac{1}{2}M + \kappa \frac{C(\bm{\Omega})}{R},
\end{align}
is diagonalized at each value of the hyperradius.
To find eigenstates of Eq.~\eqref{eq_Had},
the channel functions are expanded at a fixed hyperradius
using the antisymmetrized hyperspherical harmonics,
\begin{align}
  \Phi_{\chi}^{(M)}(R;\bm{\Omega}) = \sum_{Ka} c_{Ka}(R)\Phi_{Ka}^{(M)}(\bm{\Omega}).
\end{align}
Under this expansion,
the matrix elements of $H_{\rm ad}$ are
\begin{align}
  \label{eq_matelements}
  \langle H_{\rm ad} \rangle = U_K^{(M)}(R) \delta_{KK'}
  + \kappa \frac{\langle K'{a'}| C(\bm{\Omega}) | K {a}\rangle}{R}.
\end{align}
The brackets indicate that the integrals are taken only over the
hyperangles
and the ${a}$'s that enumerate the different hyperspherical harmonics
of a given $K$ manifold are distinguished by a subscript.

\begin{figure}
 \centering
  \includegraphics[angle=0,width=0.45\textwidth]{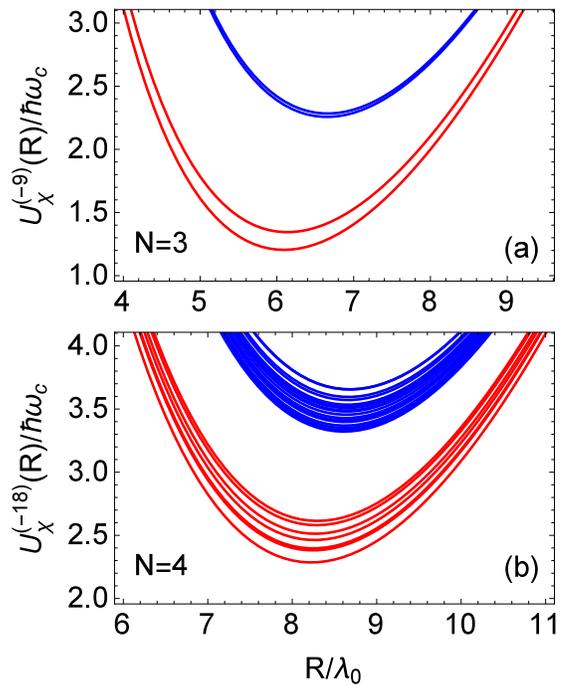} 
  \caption{(Color online) 
    Top: Adiabatic potentials $U_{\chi}^{(-9)}(R)$ for $N=3$, $M=-9$ and $\kappa=1$.
     Bottom: Adiabatic potentials $U_{\chi}^{(-18)}(R)$ for $N=4$, $M=-18$ and $\kappa = 1$.
    Similar to Fig.~\ref{fig_pot},
    the larger gaps (on the order of $\hbar \omega_c$) indicate magnetic excitations, 
    while the smaller splittings are due to Coulomb interactions. 
    The separate clusters of curves (distinguished by color online) indicate 
    different $K$ manifolds, with $K = |M|$ for the lowest grouping of curves
    and $K$ increasing from bottom to top. 
  }
  \label{fig_weak}
\end{figure}

The solutions at each $R$
describe a set of adiabatic potentials $U_{\chi}^{(M)}(R)$,
where
\begin{align}
  H_{\rm ad}\Phi_{\chi}^{(M)}(R;\bm{\Omega})
  =U_{\chi}^{(M)}(R)\Phi_{\chi}^{(M)}(R;\bm{\Omega}).
\end{align}
Fig.~\ref{fig_weak}(a) 
shows the adiabatic potentials $U_{\chi}^{(M)}(R)$ 
for $N=3$, $M=-9$, and  $\kappa=1$, 
while Fig~\ref{fig_weak}(b)
shows the adiabatic potentials for $N=4$, $M=-18$, and $\kappa=1$.
The interactions are weak such that
the different K manifolds are still distinguishable.
However,
the Coulomb interaction has split the degeneracy of the potentials.
If $\kappa$ were increased,
then the states comprising different $K$ manifolds 
would begin to overlap.

Without approximations, 
the many-dimensional Schr\"odinger equation
with Coulomb interactions
reduces to an infinite set of one-dimensional
coupled ordinary differential equations in terms of the adiabatic potentials:
\begin{align}
  \label{eq_HNIc}
  & \left\{ -\frac{1}{2\mu} \frac{d^2}{dR^2}
  + U_{\chi}^{(M)}(R) - E \right\} F_{E\chi}^{(M)}(R) \nonumber \\
  & - \frac{1}{2\mu}\sum_{\chi'}
  \left\{ 2P_{\chi\chi'}\frac{d}{dR} + Q_{\chi\chi'}\right\} F_{E\chi'}^{(M)}(R) \
  = 0.
\end{align}
The $P_{\chi\chi'}$ and $Q_{\chi\chi'}$ matrices,
\begin{align}
\label{eq_Pcoupling}
  P_{\chi \chi'}(R) = \bigg\langle \Phi_{\chi}^{(M)} \bigg| 
  \frac{\partial \Phi_{\chi'}^{(M)}}{\partial R} \bigg\rangle
\end{align}
and
\begin{align}
\label{eq_Qcoupling}
  Q_{\chi \chi'}(R) = \bigg\langle \Phi_{\chi}^{(M)} \bigg| 
  \frac{\partial^2 \Phi_{\chi'}^{(M)}}{\partial R^2} 
  \bigg\rangle,
\end{align} 
describe the non-adiabatic coupling between different channel functions,
the $\Phi_\chi^{(M)}$.
 
A good approximation to the adiabatic potentials is
to neglect the coupling between different $K$ manifolds
and apply degenerate perturbation theory.
Diagonalizing the Coulomb matrix in each manifold,
with restricted matrix elements $\langle Ka'|C(\bm{\Omega})|Ka\rangle$,
yields the eigenvalues $C_{K\gamma}^{(M)}$, where the subscript $\gamma$
labels the different eigenstates of the Coulomb matrix
in a fixed $K$ manifold.
The Coulomb interaction under this approximation 
reduces the problem to a set of uncoupled one-dimensional potentials,
just like in the noninteracting system.
The resulting adiabatic potentials are
\begin{align}
  \label{eq_Uapprox}
  U_{\chi}^{(M)}(R) \approx \delta_{KK'} \left( U_K^{(M)}(R) 
  + \kappa \frac{ C_{K\gamma}^{(M)} }{R} \right).
\end{align}

\begin{figure}
 \centering
  \includegraphics[angle=0,width=0.45\textwidth]{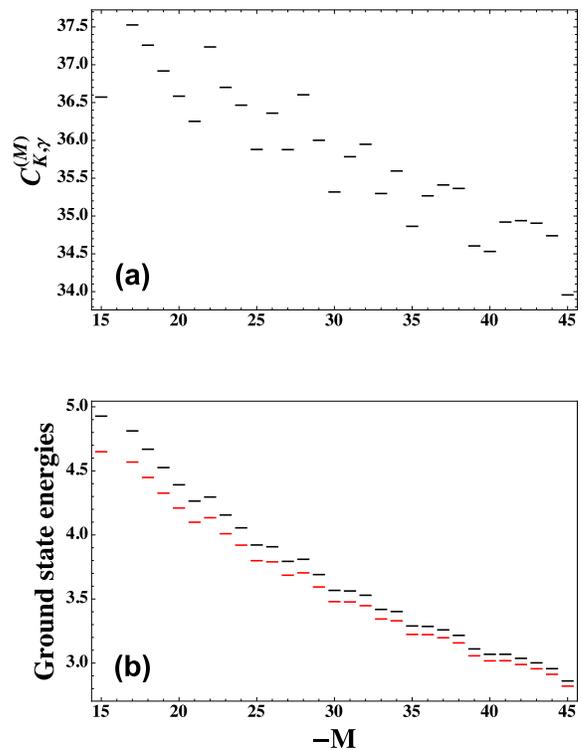} 
  \caption{(Color online) 
   (a) The minimum hyperangular Coulomb eigenvalues, the dimensionless $C^{(M)}_{K \gamma}$
    in Eq.~\eqref{eq_Uapprox} for $N = 6$ particles, 
   obtained by diagonalizing the Coulomb interaction within 
   the degenerate manifolds having $K = -M$ (which corresponds to the lowest Landau level). 
   Coulomb eigenvalues are shown for the $K = 15, \ldots, 45$ manifolds, which correspond to filling factors $\nu = 1 \ldots 1/3$. 
   (b) Two Yrast plots, plots of the minimum energy at each $M$, of the $K,M$ manifolds for six particles in the lowest Landau level using different approximations. The energies of the upper (black) curve are calculated entirely in first order perturbation theory in the lowest Landau level. For the lower energies (red), the hyperangular energies are calculated perturbatively (taken from (a) above), but the hyperradial energies are calculated using exact numerical techniques. The lower yrast spectrum constitutes a departure from the strict lowest Landau level approximation. In this model, no confinement potential is included, and particle confinement occurs only due to angular momentum conservation.}
  \label{fig_lowest6C}
\end{figure}

Fig.~\ref{fig_lowest6C}(a)
shows the lowest hyperangular Coulomb eigenvalues, the $C^{(M)}_{K \gamma}$ , for $N=6$ particles obtained by diagonalizing the Coulomb interaction within the degenerate manifolds having $K=-M$. The minimum eigenvalues are shown from the $K=15$ manifold (filling factor $\nu=1$) to the $K=45$ ($\nu=1/3$).  The $C^{(M)}_{K \gamma}$ values shown are dimensionless quantities.  A classical minimization of the Coulomb potential at fixed hyperradius gives a lower bound to the quantum eigenvalues, and the following formula gives approximate minimum values, namely:
\begin{align}
  C_{\rm min}(N) \approx (0.12+0.33/N)N^2(N-1) \label{eq_approxC}
\end{align}
For instance, for $N=6 (8)$ particles, the direct minimization gives $C_{\rm min}=31.7824 (71.5427)$, whereas the approximate formula gives 31.5 (72.2).  The value of $C^{(-45)}_{45, \gamma=1}$ computed with hyperangular wavefunctions is within 7$\%$ of this minimum value. The classical minimum hyperangular Coulomb potential corresponds to the internal geometrical distribution of particles that minimizes the energy for any size system. As $|M|$ is increased, the increasing Hilbert space size allows the particles to better approach this ideal, energy minimizing internal geometry.

Including the potentials of Eq.~\eqref{eq_Uapprox} in Fig.~\ref{fig_weak} 
only slightly alters the potentials, 
and the changes are largest for the higher energy potential curves 
within a given $K$ manifold.
The hyperangular eigenvector corresponding to the smallest eigenvalue
is the state that minimizes the Coulomb interactions.
For example,
we find 98\% overlap with the hyperangular numerical ground state and the
hyperangular part of the Laughlin function
for the three-body system with $K=-M=9$,
indicating that the ground state in this system is a quantum liquid. 
The energies the $K,M$ manifold restricted calculation can be found by multiplying the hyperangular eigenvalues, $C^{(M)}_{K \gamma}$, times the hyperradials energies, which can be found in different ways.   
Ignoring the changes to the potential curves due to the Coulomb interactions constitutes treating the hyperradial energies to first order in perturbation theory.

\begin{table*}
\caption{Comparisons to conventional systems. All rows are estimations of the ground state energy change due to the introduction of the Coulomb interaction,converted to our system of magnetic units, defined in the text. Row 1:Extrapolations of $\Delta E$ from Haldane sphere configuration interaction calculations, using quartic, least squares fits following the method in Ref.~\cite{Wooten2013}. The extrapolations become less accurate as the computational size of the systems limit the number of different sphere sizes that can be calculated. Row 2: Planar exact numerical diagonalization $\Delta E$ in the lowest Landau level: first two values from Ref.~\cite{Laughlin1983b}$*(3/\sqrt{2})$, and last three values from Ref.~\cite{Jeon2006}. Row 3: Degenerate perturbation theory calculation using hyperspherical picture within a fixed-K manifold along with a first order perturbation treatment of the hyperradial equation. Row 4: Degenerate perturbation theory within the fixed-K manifold with an exact, nonperturbative treatment of the hyperradial differential equation. Row 5: Born-Oppenheimer approximation neglecting nonadiabatic coupling matrices P and Q with exact non-perturbative treatment of the hyperradial differential equation, which constitutes a lower bound approximation when well-converged.  Row 6: Adiabatic approximation, which includes the diagonal element of the Q matrix in the lowest potential curve, with an exact, nonperturbative solution of the hyperradial differential equation. These constitute strict upper bounds to the ground state energies. }
\begin{ruledtabular}
\begin{tabular}{l | c c c c c c}
N,$-M$ & 3,$9$ & 3,$15$ & 4,$18$ & 5,$30$ &6,$45$ \\
\hline
$\Delta E$, Haldane sphere, fit, extrapolation & 0.71656 & 0.5526 & 1.310 & 2.04 & $\approx 3$\\
$\Delta E$, Planar calculations \footnote{First two values are $(3/\sqrt{2})$times values taken from \cite{Laughlin1983b}, the remaining 3 values are taken from \cite{Jeon2006} }& 0.716527 &  0.55248 & 1.30573 & 2.02725 & 2.86015 \\
\hline
$\Delta E,$ Perturbation Theory & 0.716527 & 0.55248 & 1.30573 & 2.02725 & 2.86015 \\
$\Delta E,$ Degenerate fixed-$K$ & 0.704637 & 0.54792 & 1.28552 & 1.99742  & 2.81994\\
$\Delta E,$ Born-Oppenheimer (lower bound*) & 0.70198 & 0.54722 & 1.28086 & 1.99226\footnote{The value shown may not be a converged lower bound to all digits shown.} & -- \\
$\Delta E,$ Adiabatic (upper bound) & 0.70204 & 0.54723 & 1.28092 & 1.99230 & --
\end{tabular}
\end{ruledtabular}
\label{LCtable}
\end{table*}

In the lowest Landau level, hyperspherical energy calculations using first order perturbation theory match the results of conventional planar configuration interaction calculations, which are considered the numerical standard in quantum Hall studies. In the hyperspherical treatment, perturbation theory calculations consists of multiplying the hyperangular Coulomb eigenvalues, $C^{(M)}_{K \gamma}$, times the first order perturbation calculations of the hyperradial expectation value of $1/r$ from Eq.~\ref{eq_Uapprox}. The upper (black) Yrast plot in Fig.~\ref{fig_weak}(b) for a six particle hyperspherical system under these approximations is identical to the exact numerical diagonalization plot in Fig.3 of Ref.~\cite{Jeon2006} to within the numerical accuracy shown in their Table III~\cite{YrastNote}. We present a few numerical examples for selected systems in the third row of Table~\ref{LCtable}. These values are identical to within published numerical accuracy to standard planar numerical calculations~\cite{Laughlin1983b,Jeon2006}, given on row 2, and are significantly more accurate than calculations performed using a spherical geometry, given in row 1. Numerical studies in the spherical geometry of Ref.~\cite{Haldane1983} are another standard in the quantum Hall system, but while calculations on the Haldane sphere capture much of the physics of the quantum Hall system, they are not considered numerically accurate due to finite size effects. The energy shifts given in row 1 are extrapolations from the spherical geometry to an infinite plane, performed using least squares quartic fits to the energy shifts following Ref.~\cite{Wooten2013}. Despite the numerical inaccuracy of the spherical geometry, we present these energy shifts as another comparison to standard techniques in the quantum Hall field.

The calculations can be improved from the lowest Landau level restriction by treating the hyperradial energies using exact numerical techniques. In this approximation, we maintain the single $K,M$ manifold degenerate perturbation theory approximation, but consider how the resulting Coulomb hyperangular eigenvalues alter the hyperradial potential curves. Solving the hyperradial Schr\"odinger equation using exact numerical techniques yields a slightly lower energy than the energies given using pure perturbation theory, as shown in both Fig.~\ref{fig_weak}(b) and row 4 of Table~\ref{LCtable}.  This technique constitutes incorporating additional functional space from higher Landau levels into the Hilbert space of the problem. This approximation would be analogous to a variational energy minimization based on a scaling parameter multiplying all of the single particle radial coordinates from the original Slater determinant basis. The energies are lowered in this approximation because the introduction of Coulomb repulsion expands the hyperradial potential curves outward, and the resulting hyperradial energies are lowered by this expansion.  The solutions are still confined by the diamagnetic term of the Hamiltonian, but no additional confinement potential has been included to compress the electrons into a smaller area. Including a confinement potential to model the inward Coulomb pressure due to the bulk might be appropriate in modeling many body condensed matter systems, but it is less appropriate for calculations in trapped atom systems or quantum dots.

More accurate calculations require extending beyond the single $K,M$ manifold approximation, and are not trivial. However, upper and lower bounds on the exact energies of the system can be established using additional approximations. A lower bound can be established using the Born-Oppenheimer approximation \cite{Starace1979}. In this approximation, higher $K$ manifolds with the same $M$ are included in the hyperangular energy calculations, and the hyperradial energies are calculated by solving Eq.~\ref{eq_HNIc} numerically while neglecting the nonadiabatic coupling $P$ and $Q$ matrices. If the fixed-$R$ hyperangular calculation is fully converged, these values can be proven to be lower bounds on the ground state energies for each of the corresponding symmetries. A few example lower bounds are given in row 5 of Table~\ref{LCtable}. The upper bounds given shown in row 6 of the same table are calculated under the adiabatic approximation, which differs from the Born-Oppenheimer approximation by including the diagonal elements of the $Q$ matrix in the numerical solution to Eq.~\ref{eq_HNIc}. When fully converged, these values give strict upper bounds to the ground state energy for those symmetries. As before, these upper and lower bounds apply to a system with no additional confining potential besides the diamagnetic confinement. 

The fact that the upper and lower bounds in this hyperspherical representation 
differ by less than $10^{-4}$ for all cases shown is strong evidence that the adiabatic representation is unusually effective in this system. By comparison, the ground state energies of the hydrogen negative ion 
were found to differ by 1.7\% in the corresponding upper and lower bound levels of the approximation, 
(see \cite{Klar1978}). A major reason why the adiabatic formulation is much more accurate for the quantum Hall problem is that the charged particles here do not experience any attractive potentials, which cause potential valleys in the H$^-$ system that have much stronger nonadiabaticity.

\begin{figure}
 \centering
  \includegraphics[angle=0,width=0.45\textwidth]{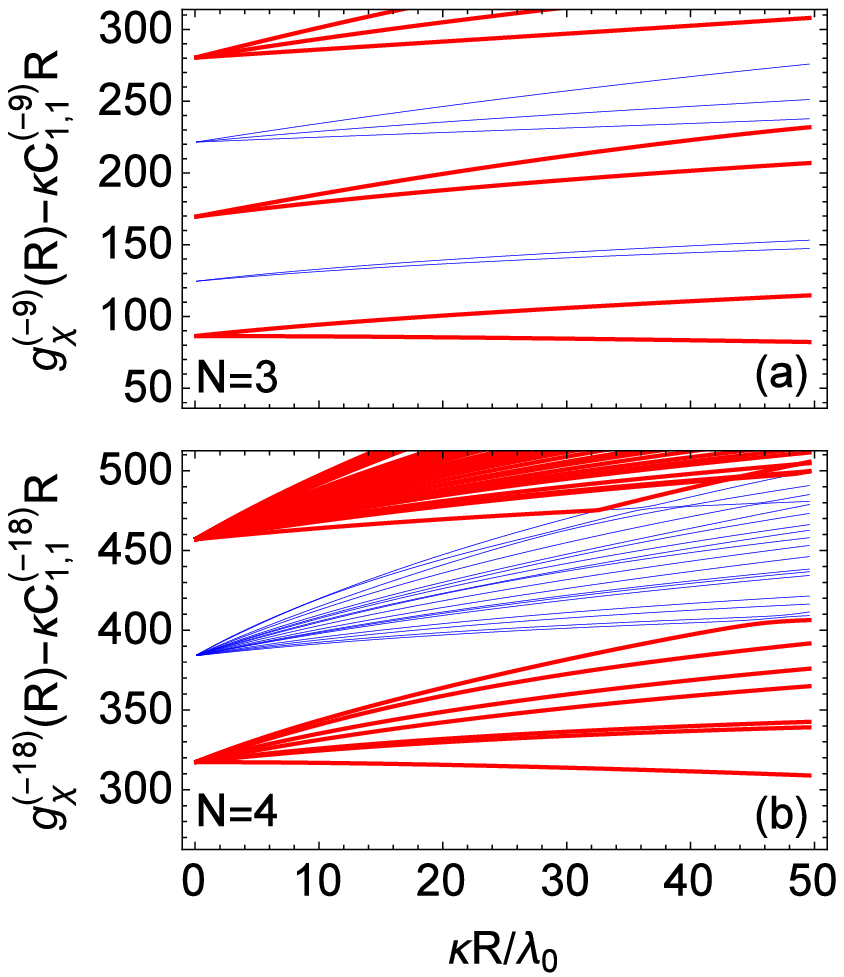} 
  \caption{(Color online) 
    Top: $N=3$ scaled adiabatic potentials $g_{\chi}(R)$ [see Eq.~\eqref{eq_gR}]
    shifted by the smallest eigenvalue $C_{1,1}^9$ from diagonalizing
    the Coulomb interaction in the $K=-M=9$ manifold.
    Bottom: $N = 4$ scaled adiabatic potentials $g_{\chi}(R)$ 
    shifted by the smallest eigenvalue $C_{1,1}^{18}$ from diagonalizing 
    the Coulomb interaction in the $K = |M| = 18$ manifold. 
    Alternating thick (red) and thin (blue) lines label different $K$ manifolds.
  }
  \label{fig_LLmixing}
\end{figure}

Another indication that the adiabatic hyperspherical representation is particularly appropriate is that the adiabatic potential curves show only weak nonadiabatic coupling between different manifolds, as evidenced by a relative lack of avoided crossings.
The weak coupling between different manifolds makes it difficult to discern any avoided
crossings in Fig.~\ref{fig_weak} on the scale shown.
To better visualize any avoided crossings,
Fig.~\ref{fig_LLmixing}
shows the scaled adiabatic potentials $g_{\chi}^{(M)}(R)$,
\begin{align}
  \label{eq_gR}
  g_{\chi}^{(M)}(R)=R^2 \left( U_{\chi}^{(M)}(R) - [(\mu/8)R^2 + M/2]\right),
\end{align}
for the same three- and four-body systems as in Fig.~\ref{fig_weak}
($M=-9$ and $M = -18$, respectively, each with $\kappa=1$),
though on a much larger scale in $R$.
At $R=0$,
the system reduces to the eigenvalues of the $\hat{\bm{K}}^2$ operator
[see Eq.~\eqref{eq_UNI}],
while at small $R$ the $g_{\chi}(R)$ are linear in $R$ with
slopes given by $\kappa C_{K\gamma}^{(M)}$.
All of the $C_{K\gamma}^{(M)}$ are positive,
so the curves in Fig.~\ref{fig_LLmixing}
have been shifted by the smallest eigenvalue
of the lowest $K$ manifold to put the curves within the same scale.

With this scaling,
many close avoided crossings become visible through the higher $K$ manifolds.
In general,
most of the crossings appear diabatic in nature.
These crossings occur at much larger hyperradii (here, $\kappa R \gtrsim 50\lambda_0$)
than the scale of Fig.~\ref{fig_weak} ($R < 11\lambda_0$). 
For comparison, 
99\% of the noninteracting hyperradial wave function is contained within $R < 8.35 \lambda_0$ for 3 particles
and $R < 10.25 \lambda_0$ for 4 particles.
Even well outside of this region, the lowest curve of the lowest $K$ manifold
remains isolated from the rest, which suggests the adiabatic approximation is a good approximation
for the ground state of the lowest Landau level.
In addition, although Fig.~\ref{fig_LLmixing} has $\kappa = 1$, the curves are universal, and their behavior changes smoothly with a trivial scaling in $\kappa$. In other words, the excellence of the adiabatic approximation applies within the relevant potential range for a very wide range of magnetic fields.

\begin{figure}
 \centering
  \includegraphics[angle=0,width=0.45\textwidth]{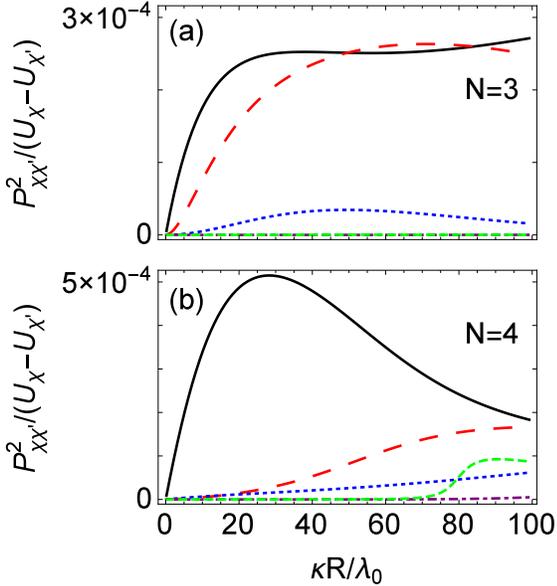} 
  \caption{(Color online) 
    $P_{\chi \chi'}^2(R)/[U_{\chi}^{(M)}(R)-U_{\chi'}^{(M)}(R)]$ as a function of $\kappa R$
    for $N=3$ and $M=-9$ (top) and $N=4$ and $M = -18$ (bottom) for $\chi = 1$.
    Solid, long-dashed, dotted, dash-dotted, and short-dashed lines are for $\chi'=2, 3, 4, 5$ and 6,
    respectively.
  }
  \label{fig_LLmixing2}
\end{figure}

A quantitative measure of the non-adiabatic coupling strength
is the dimensionless quantity $P_{\chi \chi'}^2(R)/[U_{\chi}^{(M)}(R)-U_{\chi'}^{(M)}(R)]$.
Figure~\ref{fig_LLmixing2}
shows this quantity as a function of $\kappa R$
for $N=3$ and $M=-9$,
and $N=4$ and $M = -18$ for $\chi = 1$
and $\chi'=2,\ldots,6$, that is,
the five curves represent the coupling strength from the Laughlin 1/3
potential (the lowest channel) to the five next lowest channels.
Like the potentials shown in Fig.~\ref{fig_LLmixing},
these curves are also universal in the sense that they scale simply as a function of $\kappa$.
Even the strongest nonadiabatic coupling is very small compared to unity, and the coupling to other channels is even weaker still, 
indicating the validity of the adiabatic approximation.
Other higher $\chi'$ are not shown
as their coupling strength is weaker than ${\cal O}(10^{-5})$.

To conclude this section,
the Coulomb interaction acts to split the states within
a given $K$ manifold,
yet does not lead to strong nonadiabatic coupling within the region where the potentials
are deepest.
Even with Coulomb interactions,
the separation of hyperradial and hyperangular degrees of freedom
is an excellent approximation.

\section{Exceptional degeneracy\label{sectionV}}
One of the benefits in describing this system in hyperspherical coordinates
is that the set of antisymmetrized hyperspherical harmonic basis functions 
in any $K, M$ manifold forms a complete basis in the absence of interactions. 
From perturbation theory,
it is well known that,
in a set of functions,
turning on interactions will typically act to lower the energy
of the ground state relative to all the higher energy states.
This effect is strengthened by the presence of additional degeneracy
in the system.
If the basis functions prior to turning on interactions are degenerate,
then increased degeneracy in the noninteracting picture should lead
to an increased energy separation of the ground state.
In other words, $K, M$ manifolds with enhanced degeneracy
relative to their neighboring $K, M'$ manifolds
should exhibit more strongly gapped ground states.
As a result, we predict that manifolds with exceptionally high degeneracy
are likely to also be identifiable quantum Hall states.

\subsection{Exceptional degeneracy derivation}
\label{sec_excdeg}
The following 
details how the exceptional degeneracy is derived
starting from group theory.
Only the lowest Landau level is considered in this paper,
that is,
only those states with $K=-M$.

%% Exceptional degeneracy is defined as a larger degeneracy
%% of antisymmetric states
%% at a given $|M|$ relative to the growth
%% in the overall degeneracy as a function of $|M|$.
%% For a fixed number of particles $N$,
%% the average growth in the number of antisymmetric states
%% grows polynomially as $|M|^{N-2}$ for large $|M|$.

\begin{figure}
 \centering
  \includegraphics[angle=0,width=0.45\textwidth]{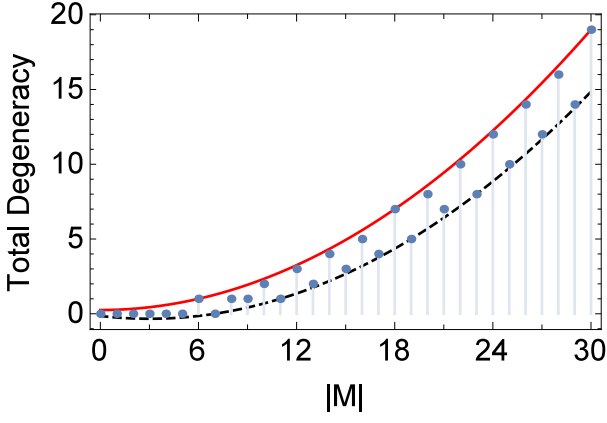} 
  \caption{(Color online) 
    Total degeneracy of antisymmetric states
    for the four-body system
    in the lowest Landau level as a function of $|M|$.
    Solid and dashed lines
    show upper and lower envelope functions, respectively, while the 
    points show the number of degenerate antisymmetric states in that
    manifold for each value of $|M| = K$.
  }
  \label{fig_N4totaldegen}
\end{figure}

First,
we derive the discrete function of $|M|$ that describes the growth
in the number of antisymmetric states.
These integer sequences are intimately related to generating functions.
For example,
from combinatorial considerations, 
the generating function $G_N(x)$
for the overall degeneracy for a fixed $N$ of spin polarized fermions
in the lowest Landau level can be derived using integer partitions.
In the lowest Landau level, each unsymmetrized $K, M$ manifold
of the relative hyperangular functions times $R^K$ forms a complete, 
translationally invariant basis of polynomials in $N = N_{rel}$ variables ($\rho*e^{i \phi}$)
that are homogeneous of order $K = |M|$.
According to Eq. (25) of~\cite{Simon2007}, the degeneracy of
the symmetric irreducible representation of this basis is equal to
\begin{align}
\label{eq_simon}
d_{sym}(K,N) = p_N(K) - p_N(K-1),
\end{align}
where $p_N(k)$ is the number of partitions of the integer $K$
into parts no longer than $N$.
The number of partitions can be calculated using a generating function~\cite{NISThandbook},
\begin{align}
\label{eq_gengenfct}
Z_N(x) = \prod_{j=1}^N \frac{1}{1-x^j} = \sum_{K = 1}^\infty x^K p_N(K).
\end{align}
Combining ~\eqref{eq_simon} and ~\eqref{eq_gengenfct} with the fact that 
there is a one-to-one mapping between the symmetric irreducible representation at $K$
and the antisymmetric irreducible representation at $K+N(N-1)/2$
yields the generating function
\begin{align}
  \label{eq_genfct}
  G_N(x) = x^{N(N-1)/2}\prod_{j=2}^N \frac{1}{1-x^j}
\end{align}
for spin-polarized fermions in the lowest Landau level.
The coefficients of the Taylor sequence 
of the generating function yields the integer sequence function whose
elements $a_{|M|}^{(N)}$ are equal to the number of degenerate 
antisymmetric states in any given $K = |M|$ manifold. 
The $a_{|M|}^{(N)}$ coefficients are equivalent to $\overline{\mathcal{N}_a}$
defined in section ~\ref{sec_relativeH}.  The way this 
generating function is used is to expand the above product in powers of 
$x$,
namely
\begin{align}
  \label{eq_sequence}
  G_N(x) = \sum_{|M|=0}^{\infty} a_{|M|}^{(N)} x^{|M|}.
\end{align}
We have verified by brute force computation that the resulting integer
coefficients $ a_{|M|}^{ (N) }$ are precisely equal to the degeneracy for
many values of $N$ and $|M|$.  
The degeneracies and generating functions can alternately be Yr
derived using group theory~\cite{Curl1960}.
The points of Fig.~\ref{fig_N4totaldegen}
show the number of antisymmetric states for the four-body system
as a function of $|M|$.
The first non-zero value at $|M|=6$ is the integer quantum Hall state.
There is only the one antisymmetric hyperspherical harmonic function
with $K=-M=6$ as is expected for a closed shell in the independent 
electron picture.
Two other notable points on this scale include those
at $|M|=18$ and 30,
which corresponds to the Laughlin $1/3$ and $1/5$ states, respectively.
Otherwise,
the general trend as $|M|$ increases
is that the number of antisymmetric states oscillates about an
overall polynomial growth.

There are many approaches to quantify the small variations in degeneracy
on top of this polynomial growth,
such as comparing the degeneracy of nearest-neighbors
or making comparisons after dividing out the largest power in $|M|$.
We choose to derive two polynomial functions that envelop the 
degeneracies, and
then compare the relative heights above the lower envelope.
The top envelope function $\bar{a}_N^{(M)}$
is forced to go through the integer quantum Hall point
at $|M|_{IQH}=N(N-1)/2$, % for fermions,
%% or $|M|_{IQH}=0$ for bosons
while the bottom function $\underline{a}_N^{(M)}$
is forced to go through the zero degeneracy value
at $|M|_{IQH}+1$.
The solid and dashed lines of Fig.~\ref{fig_N4totaldegen}
show the upper and lower envelope functions, respectively
[see also Eqs.~\eqref{eq_upperenvelope} and~\eqref{eq_lowerenvelope}].

The envelope functions are derived directly from
the exact integer sequence degeneracy function $a_{|M|}^{(N)}$.
For small systems,
Mathematica can usually find the sequence functions directly
by using SeriesCoefficient[$G_N(x),\{x,0, |M|\}$]
($-M$ is assumed to be non-negative).
In our experience, it is easier to first do a partial fraction 
decomposition
of Eq.~\eqref{eq_genfct} using Mathematica's Apart[$G_N(x)$], and 
then find the series coefficient of each term and add them together.
Regardless,
the sequence function $a_{|M|}^{(N)}$ 
is grouped in powers of $|M|$. 
The coefficients of the upper (lower) envelope polynomial are derived by
evaluating the coefficients of $a_{|M|}^{(N)}$ at $|M|_{IQH}$ ($|M|_{IQH}+1$).

As a concrete example,
for four particles
the partial fraction decomposition of $G_4(x)$ is
\begin{align}
  \label{eq_4bodyexample}
  G_4(x) = &
  \frac{1/24}{(1-x)^3}
  + \frac{-1/8}{(1-x)^2}
  + \frac{23/288}{1-x}
  + \frac{1/16}{(1+x)^2}
  \nonumber \\
  &+ \frac{-5/32}{1+x}
  + \frac{-(1-x)}{8(1+x^2)}
  + \frac{2+x}{9(1+x+x^2)}.
\end{align}
Converting Eq.~\eqref{eq_4bodyexample} to $a_{|M|}^{(4)}$
(using Mathematica functions like ExpToTrig) yields
\begin{align}
  \label{eq_a4}
  &a_{|M|}^{(4)} = 
  \frac{M^2}{48}+\frac{(-1)^{|M|}-1}{16} |M|+\frac{1}{288}
  \bigg(\!\!-27 (-1)^{|M|} -1 \nonumber \\
  &+36 \sin \left(\frac{\pi  |M|}{2}\right)-36 \cos
   \left(\frac{\pi  |M|}{2}\right)+64 \cos \left(\frac{2 \pi 
     |M|}{3}\right)\!\!\bigg).
\end{align}
The upper envelope function $\bar{a}_{|M|}^{(4)}$ comes from
evaluating the coefficients of Eq.~\eqref{eq_a4} at $|M|_{\rm IQH}$.
Here $|M|_{\rm IQH}=6$, such that
the coefficient of the $|M|^2$ term is a constant 1/48 and
the coefficient of the $|M|^1$ term is 0.
The coefficient of the $|M|^0$ term is determined last.
It is found by forcing the polynomial to equal 1
when evaluated at $|M|_{\rm IQH}$,
specifically $36/48+0+x=1$, so a value of $x=1/4$ is the final term.
This yields an upper envelope function of
\begin{align}
  \label{eq_upperenvelope}
  \bar{a}_{|M|}^{(4)} = \frac{|M|^2}{48} + \frac{1}{4}.
\end{align}

The lower envelope function $\u{a}_{|M|}^{(4)}$ comes from
evaluating the coefficients of Eq.~\eqref{eq_a4} at $|M|_{\rm IQH}+1$.
Here $|M|_{\rm IQH}+1=7$, such that
the coefficient of the $|M|^2$ term is a constant 1/48 and
the coefficient of the $|M|^1$ term is $-1/8$.
The coefficient of the $|M|^0$ term is determined last.
It is found by forcing the polynomial to equal 0
when evaluated at $|M|_{\rm IQH}+1$,
specifically $49/48-7/8+x=0$, so a value of $x=-7/48$ is the final term.
This yields a lower envelope function of
\begin{align}
  \label{eq_lowerenvelope}
  \u{a}_{|M|}^{(4)} = \frac{|M|^2}{48} - \frac{|M|}{8} - \frac{7}{48}.
\end{align}

Defining the upper envelope function as $U_N(M)$ and the lower envelope function as $L_N(M)$, we have derived the following expressions in terms of the coefficient $C_N=N(N-1)/(N!)^2$ of the maximum power $|M|^{N-2}$ as:
\begin{align}
  U_3/C_3 &= |M|+3 \nonumber \\
  L_3/C_3 &= |M|-4\nonumber \\
  U_4/C_4 &= M^2+12 \nonumber \\
  L_4/C_4 &= M^2-6 |M|-7\nonumber \\  
  U_5/C_5 &= |M|^3-9 M^2+36 |M|+260\nonumber \\
  L_5/C_5 &= |M|^3-9 M^2-9 |M|-143\nonumber \\ 
  U_6/C_6 &= M^4 \! -20  |M|^3 \! +150 M^2 \! -180 |M|\!+3105\nonumber \\
  L_6/C_6 &= M^4-20 |M|^3+60 M^2+80 |M|-256\nonumber \\
  U_7/C_7 &= |M|^5-75 M^4/2+1340 |M|^3/3-825 M^2 \nonumber \\
    &-5865 |M|+328293/2\nonumber \\
  L_7/C_7 &= |M|^5-75 M^4/2+1340 |M|^3/3-2400 M^2 \nonumber \\ 
    &+6560 |M|-323576/3 \nonumber \\
  U_8/C_8 &= M^6-63 |M|^5+1400 M^4-10920 |M|^3 \nonumber \\
    &-21168 M^2+784000 |M|+5234432\nonumber \\
  L_8/C_8 &= M^6-63 |M|^5+1400 M^4-14070 |M|^3 \nonumber \\
    &+78057 M^2-552475 |M|+714734\nonumber \\
  U_{12}/C_{12} &= M^{10}-275 |M|^9+63195 M^8/2 \nonumber \\
  &-1960200 |M|^7+70516292 M^6 \nonumber  \\
  &-1440719280 |M|^5+13562493120 M^4 \nonumber \\
  &+40317868800 |M|^3-2246025672000 M^2 \nonumber \\
  &+20132954569728 |M|+289846790411904\nonumber \\
  L_{12}/C_{12} &= M^{10}-275 |M|^9+63195 M^8/2 \nonumber \\
  &-1960200|M|^7+70516292 M^6 \nonumber \\
  &-1460365830|M|^5+16263893745 M^4 \nonumber \\
  &-99340265200|M|^3+962578332375 M^2 \nonumber \\
  &-11917386104239|M|-101761341423733/2\nonumber
\end{align}

From the envelope functions,
the relative degeneracy $g_{\rm rel}$ can now be defined as
\begin{align}
  \label{eq_reldegen}
  g_{\rm rel} = \frac{a_{|M|}^{(N)} - \u{a}_{|M|}^{(N)} }
                   {\bar{a}_{|M|}^{(N)} - \u{a}_{|M|}^{(N)}},
\end{align}
that is,
the relative height of $a_{|M|}^{(N)}$ above the lower envelope,
with respect to the separation between the two envelope functions.

\subsection{Connection between degeneracy and energy gaps}
\begin{figure}
 \centering
  \includegraphics[angle=0,width=0.45\textwidth]{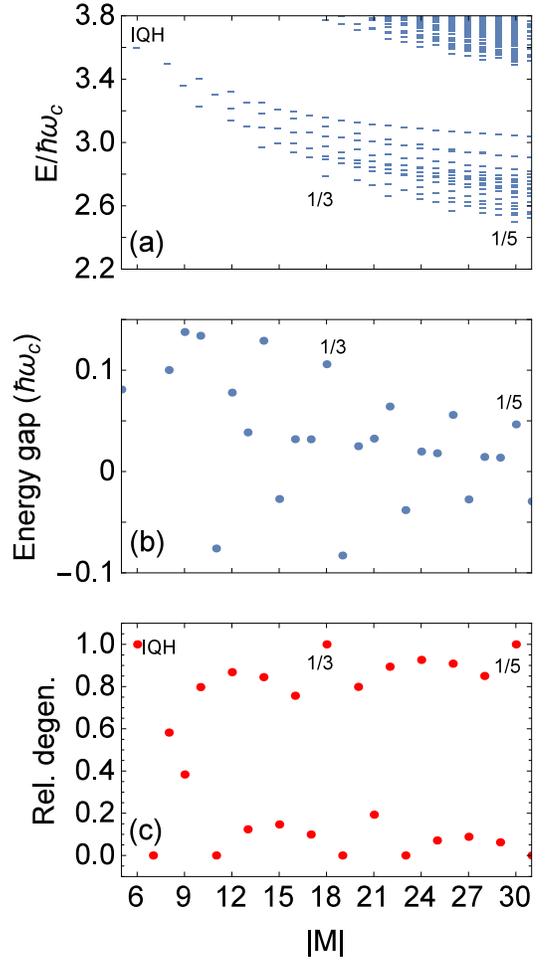} 
  \caption{(Color online) 
    Connection between exceptional degeneracy and energy gaps for $N=4$ electrons.
    (a) The dashes indicate the energies at each $|M|$.
    The IQH, $1/3$ and $1/5$ filling fractions are labeled.
    (b) The energy gap is defined as the smallest energy difference between the
    ground state at $|M|$ and all other states with $|M'| \leq |M|$.
    This could be negative if the comparative state is lower in energy.
    The IQH state's energy gap is not shown, but is about $1\hbar\omega_c$,
    since the next nearest state is approximately one vibrational quantum away.
    (c) The relative degeneracies are defined in Sec.~\ref{sec_excdeg};
    see specifically Eq.~\eqref{eq_reldegen}.
  }
  \label{fig_excdeg}
\end{figure}

Figure~\ref{fig_excdeg}
illustrates the connection between the energy gaps that appear
when solving Schr\"odinger's equation for different $M$
and the exceptional degeneracies.
Panel (a) shows the energies from solving the four-body Schr\"odinger 
equation in distinct $K, M$ manifolds
for $\kappa=1$ and various $|M|$,
but ignoring coupling between $K$ manifolds.  
The IQH state is recognizable at $|M|=6$ as it is non-degenerate
and isolated from the rest.
The Laughlin $1/3$ state at $|M|=18$ also shows a significant lowering of the energy
compared to neighboring $M$ values.
Panel (b) shows the energy gaps,
that is,
the smallest energy difference between the
ground state at $|M|$ and all other states with $|M'| \leq |M|$.
The IQH state is not shown on this scale
as its energy gap is $\approx \hbar\omega_c$.
Again, the $1/3$ state, among others,
shows a prominent positive energy gap.
Panel (c) shows the relative degeneracies.
The IQH, $1/3$, and $1/5$ states all have unity relative degeneracy,
while other states also show prominently.
Remarkably,
although the upper envelope is derived based only on the IQH state,
it passes through all Laughlin-type degeneracy points as well,
this precise matching of the upper envelope function does not hold 
for all higher particle numbers $N$.  In general, we designate those 
$K=-M$ states with relative degeneracy close to one as the exceptionally 
degenerate states, and suggest that these are candidates for observable 
$N$-body ground states that are markedly lower in energy than their 
neighbors.

\begin{figure}
  \centering
  \includegraphics[angle=0,width=0.45\textwidth]{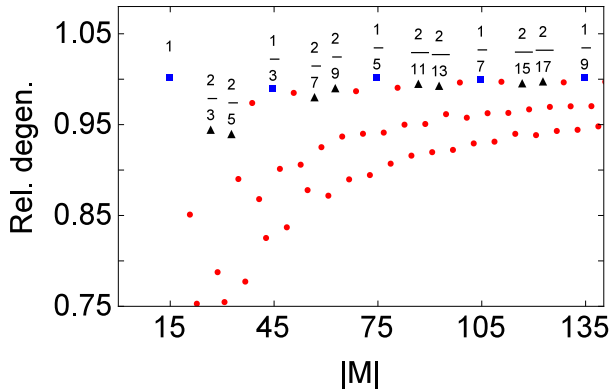} 
  \caption{(Color online)
    Relative degeneracies [see Eq.~\eqref{eq_reldegen}]
    for the six-body system are shown
    as a function of $|M|$.
    Squares show the integer quantum Hall effect and the Laughlin
    $\nu = 1/3, 1/5, \ldots$ states;
    triangles show the Jain states of two filled composite Landau levels
    (also called $\Lambda$ levels);
    circles show the remaining unidentified states.}
  \label{fig_N6reldegen}
\end{figure}

Figure~\ref{fig_N6reldegen}
shows the relative degeneracy as calculated from Eq.~\eqref{eq_reldegen}
for the six-body system as a function of $|M|$.
The squares identify the degeneracies in systems with values of $M$ 
that include the integer quantum hall and Laughlin states.
The triangles identify the degeneracies of systems that include 
the Jain states of two filled composite fermion Landau levels,
which also show a large relative degeneracy as compared to their neighbors
(which appear near $g_{\rm rel} \approx 0$)
and their next-nearest neighbors.
Circles show the remaining unidentified states.
A brief discussion of the identification of Laughlin and Jain states is given
in appendix~\ref{appendix}.
\\

\section{Conclusions\label{sectionVI}}

For testably small systems in the lowest Landau level, many of the $K,M$ manifolds of antisymmetrized functions  contain the identifiable Laughlin and Jain composite fermion states for few-body systems. The $K,M$ manifolds of the identifiable Laughlin and Jain composite fermion states all exhibit exceptionally high degeneracy compared to the majority of other $K,M$ manifolds in the lowest Landau levels. Although the relative degeneracy does not uniquely identify the known composite fermion filling states, our results suggest that high degeneracy plays a role in strengthening the energy gap of observed and described fractionally filled states. As such, it may also be interesting to examine the low-lying ground states that are associated with exceptionally high relative degeneracy, but which are not associated with the Laughlin or Jain sequences.

The adiabatic hyperspherical potential curves we have calculated are astonishingly devoid of strong couplings and avoided crossings, which is a sign that the adiabatic approximation is extremely and unusually accurate for this system in the parameter range typically probed in FQH experiments. In other words, the hyperradial degree of freedom is accurately quasi-separable in the experimentally studied range of the FQHE. Although the hyperspherical adiabatic approximation in the lowest Landau level reproduces the results of exact numerical configuration interaction calculations, the treatment accesses different aspects of the problem than either the Slater determinant construction or the composite fermion picture. The quasi-separability of the hyperradial degree of freedom is a feature not considered in other treatments of the quantum Hall system, and the hyperradius does not naturally emerge from either the Slater determinant or the composite fermion constructions of the quantum Hall system. At this time, we cannot draw any clear connections between our construction and the composite fermion framework, but suggest that this alternate perspective on the problem may allow us to examine properties of the quantum Hall system that emerge more naturally out of the hyperspherical picture. In particular, the hyperspherical construction suggests the existence of higher energy states that are hyperradial excitations of the ground state wave functions. These hyperradially excited states should have the same internal structure as their ground-state counterparts; excitations between such states should represent a breathing mode that could be probed spectroscopically.  These aspects will be explored in future publications.

\section{Acknowledgments}

We thank Gabor Csathy and Birgit Kaufmann for informative discussions. 
Critical readings of a preliminary version of this manuscript by Joe Macek and John Quinn
are also appreciated.

This work has been supported in part by a Purdue University
Research Incentive Grant from the Office of the Vice President for Research. Some numerical calculations were performed under NSF XSEDE Resource Allocation No. TG-PHY150003.

\bibliography{mybib}{}

\appendix
\section{Relative $M$ and Identification of Quantum Hall States}
\label{appendix}
\begin{table}
 \caption{Sample list of identified N-body quantum Hall states in the lowest Landau level. $M$ is the total relative azimuthal quantum number of Laughlin and Jain states identified by exact numerical diagonalization in a spherical geometry~\cite{Haldane1983}. $\nu_{CF}$ gives the filling factor of identified QH states according to the Jain composite fermion picture, including a correction that accounts for the finite size shift associated with the spherical geometry.  $\nu_{HS}$ is the calculated hyperspherical filling factor, given by Eq.\eqref{eq_fillingfactor}. The final column gives a finite size correction to the hyperspherical filling factor.}
\begin{ruledtabular}
\begin{tabular}{ c | c  c  c  c}
N & $-M$ & $\nu_{CF}$ & $\nu_{HS}$ & $(\frac{1}{\nu_{CF}}-\frac{1}{\nu_{HS}})$ \\
\hline
3 & 3 
& 1 & 1 & 0 \\
  & 9 
& $\frac{1}{3}$ & $\frac{1}{3}$ & $0$ 
\\
  & 15 
& $\frac{1}{5}$ & $\frac{1}{5}$ & $0$ 
\\
\hline
4 & 6 
& 1 & 1 
& 0 \\
& 12 & $\frac{2}{5}$ & $\frac{1}{2}$ & $-\frac{1}{2}$ 
\\
    & 18 & $\frac{1}{3}$ 
& $\frac{1}{3}$ & $0$ \\
    & 24 & $\frac{2}{7}$ 
& $\frac{1}{4}$ & $-\frac{1}{2}$ \\
    & 30 & $\frac{1}{5}$ 
& $\frac{1}{5}$ & $0$ \\

\hline
6 & 15 
& 1 & 1 & 0 \\
      & 27 & $\frac{2}{3}$ 
& $\frac{5}{9}$ & $-\frac{3}{10}$ \\
& 33 
& $\frac{2}{5}$ & $\frac{5}{11}$ & $  \frac{3}{10}$ 
\\
    & 45 & $\frac{1}{3}$ 
& $\frac{1}{3}$ & $0$ \\
    & 57 & $\frac{2}{7}$ 
& $\frac{5}{19}$ & $-\frac{3}{10}$ \\
    & 75 & $\frac{1}{5}$ 
& $\frac{1}{5}$ & $0$ \\
\end{tabular}
\end{ruledtabular}
\label{ffractable}
\end{table}

\begin{table}
\caption{Jain Composite Fermion states in the lowest Landau level.  The $M$ for an $N$-particle system at filling fraction $\nu = \nu^*/(1+2p\nu^*)$ are given for the composite fermion states most strongly observed in experiments, where $p$ is an integer indicating the number of pairs of composite fermion flux tubes attached to each electron.}
\begin{ruledtabular}
\begin{tabular}{ c | l | l}
$\nu^*$  & $-M$ & Restrictions \\
\hline
1 & $\frac{N(N-1)}{2} (2p+1)$ & \\

2 & $N(\frac{N-4}{4} + p(N-1))$ & even N only \\

-2 & $N(-\frac{N-4}{4} + p(N-1))$ & even N only \\

3 & $N(\frac{N-9}{6} + p(N-1))$ & $N\bmod{3} = 0$ only \\

-3 & $N(-\frac{N-9}{6} + p(N-1))$ & $N\bmod{3} = 0$ only \\
\end{tabular}
\end{ruledtabular}
\label{CFtable}
\end{table}

The identification of the experimentally observed fractional quantum Hall states in systems with a modest number of particles is not trivial. Although the high exceptional degeneracy of a $K,|M|$ manifold is highly correlated with the presence of a quantum Hall ground state, it is not demonstrated to be a diagnostic of the presence of a quantum Hall state. In addition, the filling fraction as given by Eq. \eqref{eq_fillingfraction} is correct in the thermodynamic limit but is only approximately correct for small systems. It is also of limited use for uniquely identifying the quantum Hall ground states. Instead, the fractional quantum Hall states of important filling factors are identified by using results from conventional, exact numerical diagonalizations in finite systems using planar, spherical, or toroidal geometry. 

For example, in a system of 6 particles, Eq. \eqref{eq_fillingfraction} would predict that the $\nu = 1/3$ state should appear when the single particle Hilbert space is restricted to 18 orbitals in the lowest Landau level, or, in other words, when the number of magnetic flux quanta in the system, $N_\phi = BA/\phi_0$, is 18. This would correspond to a planar system with $m$ restricted to $m_i = 0, -1, \ldots, -17$.   However, traditional numerical diagonalization identify the highly-gapped $\nu = 1/3$ state in a slightly smaller system where $N_\phi = 16$ and $m_i = 0, -1, \ldots, -15$.  The numerical ground state is a state with $M = -45$ and exhibits the signature of a quantum Hall state in numerical trials: a non-degenerate, translation and rotation invariant ground state with a strong energy gap. This numerical ground state is nearly identical to the famous Laughlin \textit{ansatz} wave function for many different numbers of particles~\cite{Laughlin1983,Laughlin1983a, Fano1986, Prange1987} and has been identified as the ground state of the $\nu = 1/3$ system.

The small correction to the filling factor calculated using Eq. \eqref{eq_fillingfraction} is due to the finite size of the system, and the uncorrected filling factor approaches the ideal rational fractions of the experimental system in the thermodynamic limit. The precise locations of many quantum Hall states have been established in numerical trials for a wide variety of states. The $M$ of the Laughlin filling functions ($\nu  = 1/m$ for $m = $ odd integers) are easy to establish based on the form of the Laughlin wave function. For a system with $N$ particles, the Laughlin $1/m$ wave function on the plane always occurs at $M = -mN(N-1)/2$. The relative azimuthal angular momentum, $M$, of in the independent particle picture is always a good quantum number, and is the same as the $M$ of the hyperspherical picture. As a result, we use the conventional system to identify which $M$ manifolds in the hyperspherical system contain the previously identified quantum Hall states.

The locations of the Jain composite fermion states on the plane (e.g. $\nu = 2/5, 3/7, etc \ldots$) were established by using the Jain composite fermion picture~\cite{Jain1989,jainbook}. The composite fermion sequence is found for choices of $\nu^* = 1, \pm2, \pm3, \ldots$ and positive integer $p$ at the filling factors $\nu$ given by
\begin{equation}
\nu = \frac{\nu^*}{1+2p\nu^*}.
\end{equation}
The strongest composite fermion states correspond to smaller values of $|\nu^*|$ and $p$. We have used the composite fermion construction on the Haldane sphere~\cite{Haldane1983} to identify the planar $M$ values for the Jain states. Because these electronic wave functions on the sphere involve only single-particle wave functions in the lowest Landau level, they can be mapped straightforwardly from the Haldane sphere to the infinite plane according to a stereographic mapping~\cite{Fano1986}. The planar $M$ values for the strongest quantum Hall states for three, four, and six particles are shown in Table~\ref{ffractable}. The filling fractions of the composite fermion picture ($\nu_{CF}$) are corrected to their values in the thermodynamic limit. For a more general system, the values of $M$ for the strongest composite fermion states can be calculated according to Table~\ref{CFtable}.
Other hyperspherical filling factors cannot be matched to a filling factor in the thermodynamic limit in the absence of either a theoretical picture (i.e. CF theory) or a series of numerical trials with many more particles that would allow the unidentified states to be extrapolated to be extrapolated to the many-particle case.

\end{document}